\documentclass[pdflatex,sn-mathphys-num]{sn-jnl}% Math 
\usepackage{geometry}
\selectfont
\usepackage{graphicx}%
\usepackage{multirow}%
\usepackage{subcaption}
\usepackage{amsmath,amssymb,amsfonts}%
\usepackage{amsthm}%
\usepackage{mathrsfs}%
\usepackage[title]{appendix}%
\usepackage{xcolor}%
\usepackage{textcomp}%
\usepackage{manyfoot}%
\usepackage{booktabs}%
\usepackage{algorithm}%
\usepackage{algorithmicx}%
\usepackage{algpseudocode}%
\usepackage{listings}%
\usepackage{tikz}
\usepackage{enumitem}
\usepackage{placeins}
\usepackage{array}

\theoremstyle{thmstyleone}%

\theoremstyle{thmstyletwo}%

\theoremstyle{thmstylethree}%

\usepackage[most]{tcolorbox}
\usepackage{xcolor}
\usepackage[table]{xcolor}
\usepackage{tabularx}
\def\pizgg{\ensuremath{\mathrm{\pi^{0}\!\to\!\gamma\gamma}} }

\usepackage{caption}
\begin{document}

\title[Article Title]{Reconstruction of overlapping electromagnetic showers in calorimeters using Transformers}

%%=============================================================%%
%% GivenName	-> \fnm{Joergen W.}
%% Particle	-> \spfx{van der} -> surname prefix
%% FamilyName	-> \sur{Ploeg}
%% Suffix	-> \sfx{IV}
%% \author*[1,2]{\fnm{Joergen W.} \spfx{van der} \sur{Ploeg} 
%%  \sfx{IV}}\email{iauthor@gmail.com}
%%=============================================================%%

\author[1]{\fnm{Yuliia} \sur{Maidannyk}}\email{yuliia.maidannyk@cea.fr}

\author[1]{\fnm{Fabrice} \sur{Couderc}}\email{fabrice.couderc@cea.fr}

\author[1]{\fnm{Julie} \sur{Malcl\`es}}\email{julie.malcles@cea.fr}

\author[1]{\fnm{Mehmet \"Ozg\"ur} \sur{Sahin}}\email{ozgur.sahin@cea.fr}

\affil[1]{\orgdiv{IRFU}, \orgname{CEA, Université Paris-Saclay}, \orgaddress{\city{Gif-sur-Yvette}, \postcode{F-91191}, \country{France}}}

\abstract{
Accurate clustering of electromagnetic energy deposits is essential for reconstructing photons and electrons in modern hadron collider experiments, where boosted topologies and pileup often cause overlapping showers and ambiguous energy assignment. We present deep learning-based clustering approaches that reconstruct particle energy and impact position directly from calorimeter readout. The study includes a two-step strategy in which candidate seed windows are identified and then jointly processed via distance-weighted message passing or attention mechanism, and a single-step graph transformer, \mbox{ClusTEX}, which performs candidate selection and reconstruction in one inference stage. \mbox{\mbox{ClusTEX}} uses a novel positional encoding scheme that separates local coordinates within the graph from global detector coordinates, enabling efficient, geometry-aware inference.
Models are trained on \textsc{Geant4} simulations of a simplified toy calorimeter and an ECAL-inspired topology with an explicit $(\eta, \phi)$ dependence. Performance is evaluated using efficiency, energy and position resolutions, background rejection, and ``splitting rate" --- the probability to reconstruct two objects for a single-photon shower. In the toy calorimeter, attention-based interactions improve the reconstruction of overlapping showers relative to both the standard algorithm and distance-driven message passing, while maintaining performance on isolated photons and reducing splitting without multi-pass inference. In boosted $\pi^0 \to \gamma\gamma$ events, the attention-based model retains di-photon mass reconstruction capability, where the standard algorithm becomes inefficient. In the ECAL-inspired topology, \mbox{ClusTEX} provides the best overall performance, yielding improved energy resolution and reduced splitting compared to two-step approaches and the standard algorithm. \mbox{ClusTEX} also remains robust under localized detector failures, showing improved stability and partial recovery of energy in non-responsive channels.
}


\maketitle

\section{Introduction}\label{secIntro}
In modern hadron collider experiments with a large number of simultaneous collisions and highly granular detectors, reconstructing the properties of individual particles from detector information is a major computational and algorithmic challenge. Electromagnetic calorimeters record the energy deposits left by particles and play a central role in the measurement of electron and photon four-momenta. To reconstruct the energy and position of the initiating particle, energy deposits in many detector cells must be clustered together, often in the presence of overlapping showers, detector nonuniformities, and experimental noise. These aspects can be illustrated using the Electromagnetic CALorimeter (ECAL) of the Compact Muon Solenoid (CMS) detector as an example.

The CMS experiment is a general purpose detector located at the Large Hadron Collider (LHC)~\cite{cms} at CERN. Its physics programme includes precision studies of the Standard Model and searches for physics beyond it. These goals require the reliable identification and reconstruction of particles produced in proton-proton collisions at center of mass energies up to $13.6$~TeV. In the High Luminosity LHC (HL-LHC) era, the delivered dataset and the number of additional simultaneous collisions (pileup) per bunch crossing will increase significantly, motivating the Phase-2 upgrade of the CMS detector and its reconstruction algorithms to sustain performance under these conditions.

A large number of physics analyses performed with CMS data, including studies of Higgs boson properties~\cite{higgs,higgs_mass}, rely on a precise reconstruction of photons and electrons. In CMS, this reconstruction is carried out through a multi-step workflow, starting from localized energy deposits in the calorimeters and culminating in particle identification and origin assignment within the particle flow framework~\cite{CMS:2020uim}.The reconstruction of the kinematic properties of photons is implemented in the Particle Flow clustering algorithm, PFClustering~\cite{pfclustering}. While this method is robust and efficient, it has limitations that become increasingly relevant in dense environments, at low energy, or for light boosted objects decays. In particular:

\begin{enumerate}
\item The increase in pileup expected at the HL-LHC will lead to a higher occupancy in the calorimeters meaning clustering ambiguities will become more frequent and the probability of incorrect associations between deposits and initiating particles will increase.
\item Its ability to separate nearby electromagnetic showers is limited. This affects the reconstruction of close-by photons, such as those produced in neutral pion ($\pi_{0}$) decays, which can mimic the signature of an isolated photon. Similar topologies also appear in searches for new phenomena, for instance the exotic decay $H \rightarrow AA \rightarrow 4\gamma$~\cite{exotic_higgs}, where $A$ denotes a light scalar or pseudoscalar particle. In order to cope with such signatures, some searches have employed end-to-end regression networks that directly target the final state observable (for example the reconstructed mass). While effective for a specific topology, such approaches are intrinsically specialized and cannot be straightforwardly reused across other physics analyses and search channels.
\item The algorithm exhibits a high background rate at very low energies (below approximately $1$~GeV) related to detector noise. Such backgrounds are expected to increase with detector aging and can further degrade reconstruction performance.
\end{enumerate}

In addition to these aspects, the reconstruction of converted photons constitutes a particularly challenging regime. Photon conversions modify the observed topology through additional physics objects and the associated broadening of the spatial distribution of energy deposits, increasing the probability of overlaps and making the association of deposits to the correct initiating particle more difficult and ambiguous due to similar topologies associated to decays of other particles, and/or pileup. Improvements in the clustering stage can therefore directly benefit the reconstruction of converted photons, in addition to the gains obtained for unconverted photons.

To address these limitations and improve the reconstruction of particle properties, machine learning (ML) methods can be considered. ML techniques are already widely used in  analyses involving photons and electrons (for example \cite{art1,art2}) and even in the future detector design optimizations (see for example \cite{art3, art4}). However, ML has not been adopted widely for energy clustering in electromagnetic calorimeters in a way that is optimized for multiple overlapping deposits originating from several particles, where new challenges emerge from high occupancy and detector-specific effects.

Earlier work introduced the DeepCluster concept as a graph neural network (GNN) based approach to calorimeter clustering~\cite{dcluster}. One shortcoming of the DeepCluster approach was the tendency to reconstruct two clusters for a single photon energy deposit due to enriched training on samples with two close-by photons. This issue was classified as `splitting' and required multi-pass inference to re-merge these clusters into a single cluster corresponding to a single photon energy deposit. In this paper, we make a substantial step beyond the earlier work along two complementary venues. First, we present an updated DeepCluster style approach in which the conventional graph message passing is replaced by an attention mechanism. This modification retains the graph representation but enables a more adaptive exchange of information between detector cells, which improves the performance of the GNN-based approach and removes the necessity for an ad hoc multi-pass inference. Second, motivated by the practical difficulty of deploying two-step clustering networks in real event conditions and in full reconstruction chains, we introduce a new single-step graph-based transformer architecture. This single-step formulation reduces the reconstruction to one unified inference stage, while preserving the ability to model intricate dependencies through attention.

A central ingredient of the proposed novel transformer approach is a positional encoding strategy specifically designed to account for the topology of the detector readout. Instead of relying on global position encodings that are combined through additive embeddings, we introduce a novel encoding based on concatenated local position information defined relative to the node neighbourhood within the graph. This design improves the network's awareness of position-dependent detector effects, such as $(\eta,\phi)$ dependencies induced by geometry, local nonuniformities including non-responsive cells, geometric energy leakage, and structures related to electronic readout, while maintaining efficient communication between graph nodes via attention. Such a combination is not naturally achievable with the GNN-based DeepCluster formulation, where information propagation is more constrained by the message-passing scheme.

Finally, we demonstrate the performance of the proposed methods not only in simplified topologies, but also in a more realistic detector simulation that incorporates geometrical nonuniformities. In particular, we include effects such as a $\phi$-$z$ tilt and crystal alignment towards interaction point that introduces an explicit $(\eta,\phi)$ dependence in the observed patterns. Furthermore, we consider possible channel failure conditions and include non-responsive readout blocks or channels in our simulations. This provides a stringent test of robustness and allows a more representative assessment of reconstruction performance under conditions closer to those encountered in practice. The results are presented for photons and neutral pions and are compared to the performance of the PFClustering algorithm and GNN in terms of energy and position resolution as well as the identification and separation of nearby electromagnetic showers.

\section{Particle reconstruction in the CMS electromagnetic calorimeter}
\label{secECAL}

The reconstruction strategy developed in this paper is formulated in a detector-agnostic way and can be applied to a broad class of clustering problems. Throughout this work, the CMS electromagnetic calorimeter is used as the reference environment. This section summarizes the relevant detector elements and the baseline clustering approach used in CMS, which serves as the primary reference for performance comparisons. We also briefly discuss the energy correction and higher-level clustering steps used in CMS, including learning-based superclustering approaches that operate on top of PFClustering outputs.

\subsection{The CMS electromagnetic calorimeter}

The CMS electromagnetic calorimeter is a scintillating lead-tungstate calorimeter optimized for precision measurements of photons and electrons. In Run~2 and Run~3, it is divided into a barrel section covering $|\eta|<1.479$ and two endcaps extending up to $|\eta|<3.0$. The barrel is a homogeneous single layer of lead tungstate crystals, with a transverse granularity comparable to the Moli\`ere radius and a depth of 23~cm, corresponding to 25.8~radiation lengths. The front face of each barrel crystal is $2.2\times2.2$~cm$^2$, while the endcap crystals have a larger front face. The crystals are arranged in a quasi-projective geometry with respect to the nominal interaction point and are tilted by 3 degrees in the barrel to minimise energy leakage through mechanical gaps \cite{cms_tdr}.

The reconstruction of electromagnetic objects in ECAL is based on the characteristic development of electromagnetic showers~\cite{calorimetry}. When a photon or an electron enters the calorimeter, it initiates a cascade of secondary particles and scintillation light is produced and collected by photodetectors. The energy is typically distributed across multiple crystals around the entry point. The reconstruction task is to infer, from the observed energy deposits, the energy and the direction of the primary particle as precisely as possible.

The HL-LHC program introduces additional constraints that directly impact calorimetric clustering. The pileup rate is expected to increase substantially, leading to higher occupancy, more frequent overlaps between showers, and a larger fraction of ambiguous energy deposits. In CMS Phase-2, the ECAL endcaps will be replaced by the high-granularity calorimeter (HGCAL), while the barrel will continue operation with upgraded electronics and reconstruction software adapted to the high-pileup environment. These conditions motivate reconstruction approaches that can reason about neighbourhood relationships in a geometry-aware manner and that remain robust to detector nonuniformities and strong $\eta$ dependencies.

\subsection{The PFClustering algorithm}

The baseline algorithm used in CMS to reconstruct electromagnetic energy deposits at the clustering stage is the Particle-Flow clustering algorithm, PFClustering~\cite{pfclustering}. PFClustering aggregates cells around a local maximum based on energy thresholds and geometry to form a topological cluster. This design ensures high reconstruction efficiency down to low energies and reliable operation in the presence of detector noise. The details of the PFClustering algorithm are discussed in Appendix~\ref{sec:pfclusteringalgo}.

The PFClustering methodology provides good performance for isolated deposits and for low-energy reconstruction. However, performance degradation becomes more pronounced in dense environments where multiple nearby local maxima compete within the same topological region. This motivates the development of ML-based clustering methods capable of exploiting richer context while preserving robustness to detector effects and geometry.

\subsection{Energy regression and higher-level clustering}

The energy reconstructed from clustered deposits can be biased by effects such as per-channel energy thresholds, shower leakage, and local detector response variations. In CMS, these effects are mitigated through multivariate regression techniques trained on simulation. In the context of PFClustering, a boosted decision tree (BDT) regression is commonly used to correct the PFCluster energy using a set of observables describing the cluster energy distribution and shape~\cite{BDT}. Typical inputs include the raw clustered energy, the reconstructed position, local energy maxima, asymmetries around the seed, and simple shape descriptors derived from energy-weighted moments. These corrections provide a strong baseline for energy resolution comparisons in this work.

Beyond PFClusters, CMS also employs higher-level clustering steps to build electromagnetic superclusters, particularly for photons and electrons whose showers can spread across a wider region due to conversions and bremsstrahlung in the tracker material. An ML-based approach in this direction, Deep Superclustering, has been proposed and studied, where the algorithm takes PFClustering outputs as inputs and learns to aggregate them into superclusters~\cite{Valsecchi_2023}. This strategy targets improvements in the reconstruction of extended electromagnetic topologies by operating on a representation that is already shaped by PFClustering. A detailed discussion of the superclustering strategy and its deep learning-based extensions can be found in Ref.~\cite{simkina:tel-04412128}.

The methodology presented in this paper targets the clustering of calorimeter energy deposits at an earlier stage than superclustering approaches that operate on PFClusters. By improving the association of deposits to initiating particles, especially in dense and conversion-rich topologies, the proposed reconstruction is expected to provide higher-quality building blocks for subsequent steps of the reconstruction workflow. In particular, improvements at the clustering stage naturally propagate to higher-level reconstruction, including regression-based energy corrections and superclustering algorithms that aggregate PFClusters into electromagnetic objects~\cite{Valsecchi_2023,simkina:tel-04412128}. The PFClustering-based chain, together with its regression and superclustering components, therefore remains a well-defined and widely used reference against which the proposed strategies are evaluated. More importantly, the proposed methodology can be considered as the first step towards a unified clustering - superclustering approach, which can be explored in future studies.

\section{Simulation and dataset}
\label{section:dataset}

To train the models presented in this paper and to compare their performance to the one of the PFClustering algorithm, dedicated calorimeter simulations are produced using \verb|Geant4 11.2|~\cite{geant}. Two independent simulation configurations are considered. The first one is a simplified electromagnetic calorimeter, referred to as a \textit{toy calorimeter}, which enables controlled studies of clustering performance in an idealized geometry. The second configuration is designed to emulate a more realistic segment of the CMS ECAL barrel by incorporating the crystal layout, the ECAL mechanical
structure, the material in front of ECAL, as well as an explicit dependence on $(\eta,\phi)$ through particle incidence from the interaction point. Further details of the simulation setups are provided in Appendix~A.

\subsection{Toy calorimeter simulation}

The toy calorimeter uses a simple geometry consisting of a single layer of lead tungstate crystals arranged in a uniform grid of 51$\times$51 crystals. The crystal material and dimensions are chosen to match the CMS ECAL barrel characteristics, namely PbWO$_4$ crystals with size $2.2\times2.2\times23$~cm$^3$.
A single photon gun is directed perpendicularly toward the detector plane and generates photons with energies uniformly distributed in the range $1$ to $100$~GeV. The impact point on the calorimeter front face is randomly sampled while avoiding the outermost crystals to mitigate boundary effects and to ensure that most of the shower energy is contained within the active volume.

\subsection{ECAL-inspired simulation with realistic topology}

\begin{figure}[ht!]
    \centering
    \includegraphics[width=1.0\linewidth]{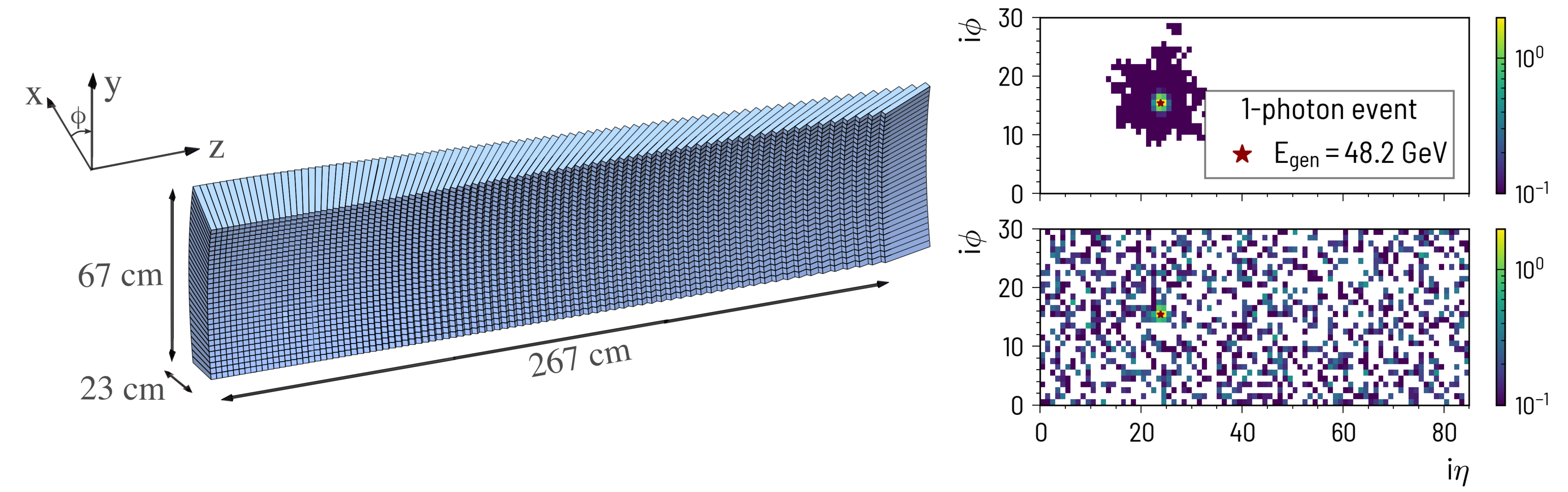}
    \caption{Left: realistic CMS ECAL simulation geometry, where each PbWO$_4$ crystal is quasi-projective and aligned toward the interaction point with a 3$^\circ$ tilt in both the $z$ and $\phi$ directions. Right (top): example calorimeter window for a single-photon event showing the energy deposits in the crystals. Right (bottom): the same calorimeter window after applying the simulated read-out noise and an energy cut at 50~MeV to mitigate contributions dominated by electronic noise}
    \label{fig:detector}
\end{figure}

A second simulation  is introduced to probe reconstruction performance under more realistic conditions. It retains the crystal geometry and the mechanical embedding of the scintillating volume, while omitting detector components that are not required for the calorimetric response in this study, such as the magnetic field and the full read-out chain. The scintillating volume consists of 85 crystals along the $z$ direction, corresponding to one side of the ECAL barrel coverage up to $|\eta|=1.479$, and 30 crystals in the radial direction, corresponding to $|\phi|<15$~degrees. The crystal segment is embedded in an aluminium alveolar module positioned at a radius of $R=1.290$~m from the interaction point. To emulate the material budget upstream of the calorimeter relevant for conversions and early shower development, an equivalent thickness of 33~mm of silicon is placed immediately in front of the calorimeter volume. Each crystal is aligned towards the interaction point with an additional 3-degree tilt in both $z$ and $\phi$ directions. The layout of the simulated detector slice together with an example event display is shown in Fig.~\ref{fig:detector}.

In this configuration, the photon gun is placed at the interaction point. The resulting showers therefore acquire an explicit dependence on $(\eta,\phi)$ due to the incident angle and the detector geometry. This provides a realistic source of position-dependent effects, including nonuniformities induced by the tilted-projective arrangement and the crystal orientation, and it enables an evaluation of reconstruction robustness in conditions that depart from perpendicular incidence.

\subsection{Datasets}

In both simulation configurations, each event records the generated photon energy and impact position, $(E_\mathrm{gen}, x_\mathrm{gen}, y_\mathrm{gen})$, together with the energy deposits in individual crystals. These deposits are stored as calorimeter windows indexed by crystal coordinates $(i\eta,i\phi)$ or the corresponding local indices in the simulated segment, and they serve as the input representation for the learning-based reconstruction models.

Two main dataset categories are constructed to provide access to isolated and overlapping shower regimes and allow systematic performance studies across the relevant topology spectrum:

\begin{itemize}
    \item \textbf{Single-photon dataset} \newline
    Each entry corresponds to a single generated photon. It represents the case of isolated electromagnetic showers and is used for training and for establishing baseline reconstruction performance in terms of energy and position resolution.

    \item \textbf{Two-photon dataset} \newline
    To model close-by electromagnetic showers, a dedicated two-photon dataset is built by superimposing two independently simulated single-photon events at the level of per-crystal deposited energies, producing composite calorimeter windows with overlapping showers. Pairs are selected such that the two impact positions are less than 3 crystals apart in the transverse plane, mimicking boosted di-photon configurations such as those arising from $\pi^{0}$ decays. Events in which both photons enter the calorimeter in the same crystal are excluded.
\end{itemize}

An additional dataset of neutral pions is used for evaluation and is introduced together with the performance results in Section~\ref{sec:results}.

\subsection{Per-crystal energy}
\label{sec:energy_resolution}

As the simulations described above do not include the full ECAL digitization and reconstruction chain, the per-crystal measured energy is obtained by smearing the true deposited energy. This emulates the combined impact of the intrinsic calorimeter resolution and the read-out electronic noise. In the case of superimposed samples, the true deposited energies are first summed crystal-by-crystal and the smearing is applied to the combined deposit.

The ECAL energy resolution can be parametrized as a quadratic sum of stochastic, noise and constant terms~\cite{cms_tdr}:
\begin{equation}
    \left(\frac{\sigma_E}{E}\right)^2    =  \left(\frac{a}{\sqrt{E}}\right)^2 +  \left(\frac{\sigma_n}{E}\right)^2  +    c^2,
\label{eq:energy_resolution}
\end{equation}
where $a$, $\sigma_n$, and $c$ denote respectively the stochastic, noise and constant contributions. For a crystal with true deposited energy $E^\mathrm{true}_\mathrm{xtal}$, the measured energy is obtained as
\begin{equation}
    E_\mathrm{xtal}  =  E^\mathrm{true}_\mathrm{xtal} \times \mathcal{N}\!\left(\mu=1;\,
    \sigma=\frac{\sigma_{E^\mathrm{true}_\mathrm{xtal}}}{E^\mathrm{true}_\mathrm{xtal}}
    \right),
\label{eq:smearing}
\end{equation}
where $\mathcal{N}(\mu;\sigma)$ denotes Gaussian smearing with mean $\mu$ and standard deviation $\sigma$, and $\sigma_{E^\mathrm{true}_\mathrm{xtal}}$ is derived from Eq.~\ref{eq:energy_resolution}.

In this study, we adopt parameters consistent with CMS ECAL barrel conditions~\cite{ecal_constant,clusterin_run3}: $a=0.03~\text{GeV}^{1/2}$, $\sigma_n=0.167~\text{GeV}$, and $c=0.0035$. This corresponds to a per-crystal variance
\begin{equation}
    \sigma^2    =    (0.03)^2 E^\mathrm{true}_\mathrm{xtal}    +    (0.0035)^2 \left(E^\mathrm{true}_\mathrm{xtal}\right)^2    +    (0.167)^2,
\label{eq:smearing_sigma}
\end{equation}
which is applied independently to each crystal. A low-energy cut at 50~MeV is applied on the smeared crystal energy to mitigate contributions dominated by electronic noise.

Overall, a total of 1.29M single-photon samples are generated with the toy calorimeter simulation and 1.25M samples with the ECAL-inspired detector simulation. The samples are split into training, validation and testing subsets with a 6:2:1 ratio. Although the event distribution is uniform across the crystal cross-section, events with photon impact positions near the crystal boundary (within ~0.1 crystal units) are expected to challenge the reconstruction algorithm. To improve the model performance in this regime, such events are artificially enhanced in the training and validation samples by increasing their fraction from 36\% to 44\%.

\subsection{Preprocessing and window construction}

The simulated calorimeter windows are sparse by construction, as the crystal transverse size is comparable to the Moli\`ere radius and electromagnetic showers remain localized around the particle entry. To construct a consistent input representation for the deep learning-based models and to mitigate sparsity, smaller fixed-size windows are extracted around candidate seed crystals.

For each event, crystals with measured energy above $0.5$~GeV are identified as seed candidates. This threshold corresponds approximately to 3 times the noise $\sigma$. Around each seed candidate, a $7\times7$ window is extracted and centered on the candidate crystal. For the purposes of training, each extracted window is labelled as either signal or background depending on whether it is centered on the crystal associated with the impact point of the generated particle (also referred to as the true seed). This procedure yields a large number of candidate windows per event and provides a realistic classification setting where true seeds must be identified among multiple local maxima.

\section{Deep learning models}
\label{secModels}

This section describes the neural networks used to reconstruct the photon kinematics, $(E_\mathrm{gen},x_\mathrm{gen},y_\mathrm{gen})$, from the $7\times7$ seed windows defined in Section~\ref{section:dataset}. In addition to the two-step architecture proposed in Ref.~\cite{dcluster}, we introduce two new model architectures:

\begin{enumerate}
    \item \textbf{Two-step models} 
    \begin{itemize} 
    \item \textbf{SeedFinder:} a classifier that assigns a signal probability to each candidate seed window.
    \item \textbf{Position and Energy Regression Network (PoEN) (two-step reconstruction):} a regression network that processes a small set of SeedFinder-selected candidate windows jointly. Two PoEN families are considered:
        {\begin{itemize} 
            \item a message-passing PoEN (GNN-based) that uses distance-driven interactions as introduced in~\cite{dcluster},
            \item an attention-based PoEN (GAT-based) with two flavours that differ only by the window encoder (convolutional vs pixel-flatten). To prevent positional ambiguity due to input invariance of the attention mechanism, a local positional encoding relative to the  highest-energy deposited seed window is included.
        \end{itemize}}
    \end{itemize}
    \item \textbf{Single-step graph transformer:} a one-step network that performs candidate selection and kinematic reconstruction once graphs are formed with the procedure of Appendix~\ref{sec:graph}. The novel local and global positional encoding strategy is introduced and defined for this model only.
\end{enumerate}

Unless stated otherwise, each node corresponds to one $7\times7$ seed window centred on a seed candidate crystal. For models that operate on multiple candidates simultaneously, the $N$ selected nodes are processed as a set.

\subsection{SeedFinder Network}
\label{sec:seedfinder}

The SeedFinder assigns each candidate window a score $P_\mathrm{SF}$ that quantifies the likelihood that the window is centred on the true seed crystal of a generated photon. Its purpose is to reduce the number of candidate windows passed to the regression stage, improving both inference cost and reconstruction stability in events that contain many local maxima.

\begin{figure}[ht!]
    \centering
    \includegraphics[width=0.9\linewidth]{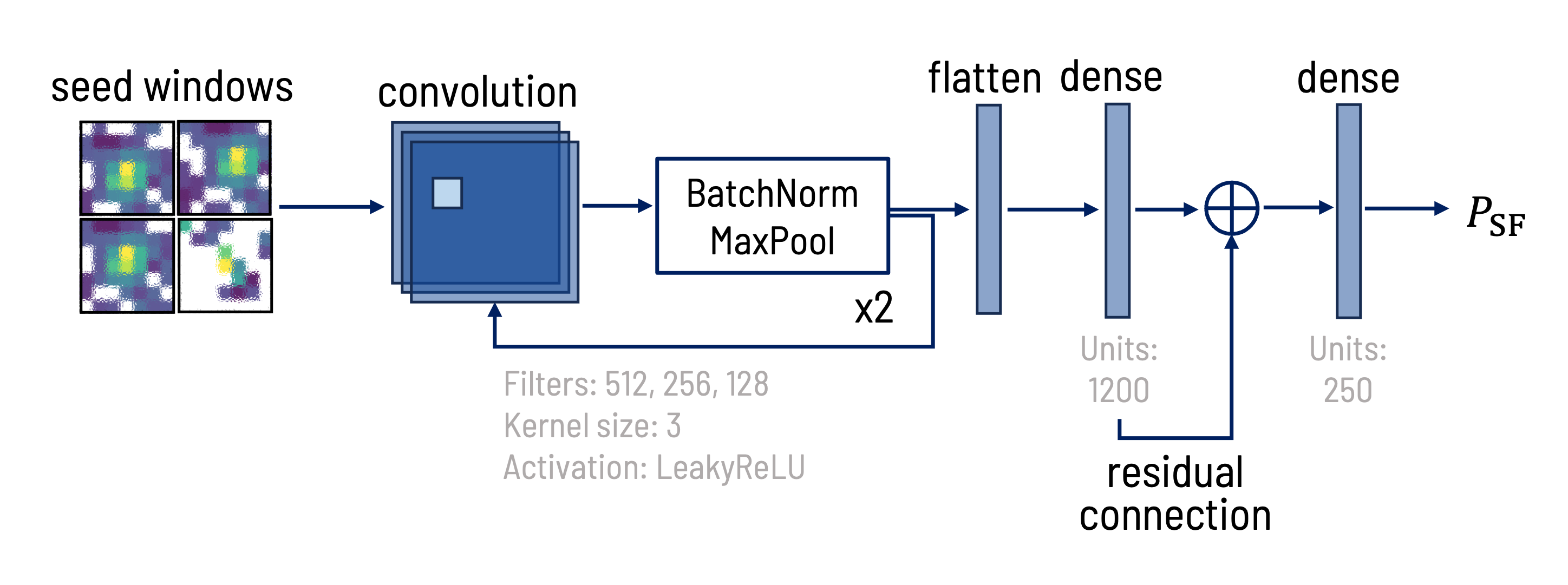}
    \caption{The optimized CNN based SeedFinder (SF) model architecture. The seed window candidates passing the threshold on deposited energy are taken as inputs to the SF model. The model assigns a probability whether the window is centred around the true seed crystal of a generated photon. The network consists of three consecutive convolutional blocks with a kernel size of three with decreasing filter sizes and three fully connected (dense) layers with residual connection in between}
    \label{fig:sf_arch}
\end{figure}

The SeedFinder is implemented as a CNN. The baseline architecture consists of two convolutional layers followed by three fully connected layers. For the ECAL-inspired simulation with realistic topology, additional architectural elements are introduced to improve robustness in the presence of geometry-driven nonuniformities. These additions include one extra convolutional layer, batch normalisation, max pooling and a residual connection in the convolutional block. The final architecture is summarised in Fig.~\ref{fig:sf_arch}.

On an event-by-event basis, candidate windows passing a threshold on $P_\mathrm{SF}$ are sorted in descending order and the top candidates are forwarded to the PoEN. For the given simulation topology, the maximum number of candidates is observed to be $N<4$ with very few occurrences of $N=4$ in cases of low threshold value on $P_\mathrm{SF}$. Therefore, we retain up to $N=4$ candidates per event. The working point on $P_\mathrm{SF}$ is chosen to optimize the trade-off between signal efficiency and background rejection. In the configuration used throughout this paper, a requirement $P_\mathrm{SF}>0.3$ yields 99.5\% efficiency for the 1-photon sample and 98.2\% efficiency for the 2-photon sample. Some of the efficiency loss is restored in PoEN due to the fact that seed windows which neighbour the true seed are considered to be background, but still carry sufficient information to reconstruct the primary photon kinematics.

\subsection{Position and Energy Regression Network (two-step reconstruction)}
\label{sec:PoEN}

The PoEN is a regression network that predicts the photon energy and impact position from a small set of $N$ candidate seed windows selected by the SeedFinder. It produces, for each candidate window $i$, an estimate
\begin{equation}
    \left(E_i^\mathrm{reco}, x_i^\mathrm{reco}, y_i^\mathrm{reco}, P_i^\mathrm{seed}\right),
\end{equation}

\noindent where $P_i^\mathrm{seed}$ is an internal seed probability used to stabilise training and to select reconstructed objects during evaluation and $(E_i^\mathrm{reco},x_i^\mathrm{reco},y_i^\mathrm{reco})$ denote the reconstructed photon energy and impact position.

A key ingredient of the PoEN is information exchange between the $N$ candidates of the same event. This is required because several candidate windows can correspond to the same generated photon, especially in dense environments or in overlapping-shower topologies. Joint processing allows the network to learn correlated candidates and reduce double-counting.

\subsubsection{Distance-weighted message-passing PoEN (DW-PoEN)}
\label{sec:dwcf}

The first PoEN variant is GNN-based and uses a message-passing style interaction between the $N$ candidates. Each seed window is first mapped to a learned embedding vector by a shared encoder. We denote the resulting node embedding by $\mathbf{h}_i\in\mathbb{R}^{d}$ for node $i$.

Node interactions are implemented with a fixed, distance-driven adjacency matrix $\mathbf{A}\in\mathbb{R}^{N\times N}$ whose entries are defined from the inverse Euclidean distance between window centres:
\begin{equation}
    A_{ij} \propto \frac{1}{\sqrt{(x_i-x_j)^2+(y_i-y_j)^2}+\epsilon},
\end{equation}
with $\epsilon=10^{-6}$ introduced for numerical stability. The message-passing update aggregates information from the other nodes using these weights to obtain updated embeddings, which are then passed through regression heads to predict $(E_i^\mathrm{reco},x_i^\mathrm{reco},y_i^\mathrm{reco})$ and $P_i^\mathrm{seed}$.

This construction enables node-to-node communication but is limited by the fact that interaction weights are fixed by a distance metric and therefore cannot adapt to node content or event context. This motivates replacing the fixed interaction pattern by attention.

\subsubsection{Attention-based PoEN (GAT-PoEN)}
\label{sec:gatcf}

A known shortcoming of the message-passing PoEN (GNN-PoEN) is its tendency to produce spurious reconstructed clusters, as reported in Ref.~\cite{dcluster}. A pragmatic mitigation based on re-processing the event in a second pass has been proposed to reduce this effect. In this work, we hypothesize that the origin of these spurious clusters is primarily related to the way information is exchanged between nearby candidates: in distance-driven message passing, close-by candidates can dominate the aggregation and propagate ambiguous features in a way that is difficult to regulate with fixed interaction weights.

To address this, the attention-based PoEN replaces the fixed distance-driven aggregation with a learnable attention mechanism that computes interaction weights from the node content. This allows the model to adaptively control how information is shared between candidates on an event-by-event basis and to down-weight neighbours that are close in space but not compatible in shower content. In this work, the attention-based PoEN is studied in two flavours that differ only by the feature extractor applied to each $7\times7$ window:

\begin{itemize}
    \item \textbf{Convolutional GAT PoEN (GAT-PoEN$_\mathrm{conv}$):} each window is processed by a convolutional encoder to produce the node embedding $\mathbf{h}_i$.
    \item \textbf{Flattened-pixel GAT PoEN (GAT-PoEN$_\mathrm{flat}$):} each window is flattened into a 49-dimensional vector and processed by a fully connected encoder to produce the node embedding $\mathbf{h}_i$.
\end{itemize}

\begin{figure}[ht!]
    \centering
    \includegraphics[width=1\linewidth]{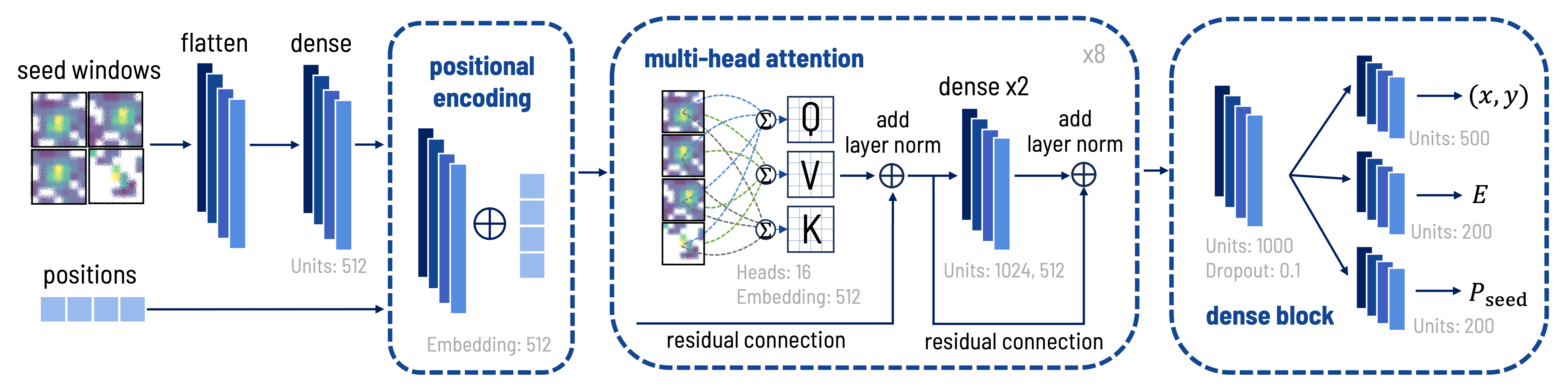}
    \caption{Architecture overview of the attention-based PoEN. The two flavours differ only by the window encoder (flattened pixels vs convolutional)}
    \label{fig:gat_arch}
\end{figure}

In both cases, the set of node embeddings $\{\mathbf{h}_i\}_{i=1}^{N}$ is passed through multi-head attention blocks. The updated embeddings are then mapped to the outputs $(E_i^\mathrm{reco},x_i^\mathrm{reco},y_i^\mathrm{reco},P_i^\mathrm{seed})$ using dedicated output heads. The architecture with the attention block is shown in Fig.~\ref{fig:gat_arch}.
A direct comparison between DW-PoEN and GAT-PoEN$_\mathrm{flat}$ is provided in Section~\ref{sec:results}.

\subsection{Single-step graph transformer with positional encoding}
\label{sec:transformer}

The two-step strategy requires a separate selection stage and a fixed choice of how many candidates are forwarded to the regression stage within a fixed window. Moreover, while it provides an excellent initial filtering stage, it faces intrinsic limitations: if the initial window is too large (for example the full CMS ECAL or layers of future high granularity detectors), a PoEN architecture needs to be adopted, while if the window is too small, edge effects appear and overlapping window choices introduce ambiguities. Furthermore, SeedFinder is inherently unaware of neighbouring seed windows, since it considers only a single window at a time.

To address these shortcomings, we introduce a single step approach, \textbf{Clustering Transformer for Energy and X-Y regression (\mbox{ClusTEX})}, in which reconstruction is performed in one inference stage, after dynamically forming an event graph around a seed candidate as described in Appendix~\ref{sec:graph}. In this formulation, candidate selection and kinematic reconstruction are learned jointly within a single model, allowing the network to decide what information to use rather than relying on a hand-tuned preselection. Forming an effective $13\times13$ window around the highest energy seed preserves locality and avoids sparsity, while the graph structure provides the flexibility to handle variable occupancy and complex topologies within the window.

\subsubsection{Positional encoding}
\label{sec:posenc}

In this work, a novel positional encoding strategy is defined for the single-step transformer as a combination of:
\begin{itemize}
    \item a \textbf{local} position defined within the effective window and concatenated into the input-window encoding,
    \item a \textbf{global} position defined with respect to detector center and summed to the node embedding, following the standard transformer formulation.
\end{itemize}

\textbf{Local position within the effective window.}
We consider one graph formed by Algorithm~\ref{alg:graph_form} and we denote by $(x_\mathrm{a},y_\mathrm{a})$ the centre of the anchor seed candidate selected in Step~1. For each node $i$ in the current graph, with window centre $(x_i,y_i)$, we define the local position relative to the anchor:
\begin{equation}
    \boldsymbol{\ell}_i =
    \left(
    \frac{x_i-x_\mathrm{a}}{d_\mathrm{effective}},
    \frac{y_i-y_\mathrm{a}}{d_\mathrm{effective}}
    \right).
    \label{eq:local_pos}
\end{equation}
This local coordinate system is shared by all nodes in the same graph and represents where each candidate lies within the effective region currently being processed.

\textbf{Window encoding and local concatenation.}
Each node corresponds to a $7\times7$ seed window and is mapped to a feature vector $\mathbf{h}_i$ by a shared window encoder. The encoder can be convolutional or based on flattened pixels, depending on the configuration. The learnable local position embedding $\boldsymbol{f}(\boldsymbol{\ell}_i)$ is then concatenated to the window features:
\begin{equation}
    \tilde{\mathbf{h}}_i = \left[\mathbf{h}_i\,;\,\boldsymbol{f}(\boldsymbol{\ell}_i)\right],
    \label{eq:concat_local}
\end{equation}
where $[\cdot\,;\cdot]$ denotes vector concatenation. 

\textbf{Global position within the detector.}
In addition to the local effective-window coordinates, each node has an absolute detector position $\mathbf{p}_i=(\eta_i,\phi_i)$. A learnable global position embedding $\mathbf{g}(\mathbf{p}_i)$ is constructed and summed to the projected node embedding:
\begin{equation}
    \mathbf{z}_i = \tilde{\mathbf{h}}_i + \mathbf{g}(\mathbf{p}_i).
    \label{eq:global_pos_sum}
\end{equation}
This term provides the model with absolute detector awareness, enabling it to learn position-dependent effects, such as $\eta$-dependent shower features and local nonuniformities.

The resulting embeddings $\{\mathbf{z}_i\}$ are then processed by attention blocks. For completeness, the attention weights within one multi-head attention layer can be written as
\begin{equation}
    \alpha_{ij} = \mathrm{softmax}_j\left(\frac{\mathbf{q}_i^\top\mathbf{k}_j}{\sqrt{d_k}}\right),
\end{equation}
with $\mathbf{q}_i=\mathbf{W}_Q\mathbf{z}_i$, $\mathbf{k}_j=\mathbf{W}_K\mathbf{z}_j$ and values $\mathbf{v}_j=\mathbf{W}_V\mathbf{z}_j$, where $\mathbf{W}_Q$, $\mathbf{W}_K$, and $\mathbf{W}_V$ are learnable projection matrices of the attention mechanism as also illustrated in Fig.~\ref{fig:gat_arch} and $d_k$ is the embedding dimension of $k_j$.
In this construction, the local information is encoded through concatenation before the input projection (Eq.~\ref{eq:concat_local}), while the global information is injected through a summed embedding (Eq.~\ref{eq:global_pos_sum}). This separation allows the network to remain sensitive to the relative structure of candidates inside the overlap window while simultaneously learning detector-position dependencies. The impact of the introduced positional encoding scheme is discussed in Appendix~\ref{sec:dead_channels_posenc}.

\subsection{Loss definition}
\label{sec:loss}
Binary cross-entropy loss is used while training the SeedFinder network. 
The loss function for PoEN and \mbox{ClusTEX} consists of three terms that correspond to the energy, position and seed probability predictions:
\begin{equation*}
\mathcal{L} = \lambda_{\mathrm{energy}}\cdot\mathcal{L}_\mathrm{energy}
            + \lambda_{\mathrm{position}}\cdot\mathcal{L}_\mathrm{position}
            + \lambda_{\mathrm{seed}}\cdot\mathcal{L}_\mathrm{seed}
\end{equation*}
with
\begin{align}
\begin{split}
\mathcal{L}_\mathrm{energy}
&= \frac{1}{N\cdot N_\mathrm{batch}}
   \sum_{i=1}^{N_\mathrm{batch}}\sum_{j=1}^{N}
   \left|E_{ij}^\mathrm{pred}-E_{ij}^\mathrm{true}\right| \\
\mathcal{L}_\mathrm{position}
&= \frac{1}{N\cdot N_\mathrm{batch}}
   \sum_{i=1}^{N_\mathrm{batch}}\sum_{j=1}^{N}
   \frac{1}{2}\left(
   \left|x_{ij}^\mathrm{pred}-x_{ij}^\mathrm{true}\right|
   + \left|y_{ij}^\mathrm{pred}-y_{ij}^\mathrm{true}\right|
   \right) \\
\mathcal{L}_\mathrm{seed}
&= -\frac{1}{N\cdot N_\mathrm{batch}}
   \sum_{i=1}^{N_\mathrm{batch}}\sum_{j=1}^{N}
   \alpha(1-P_{ij}^\mathrm{seed})^\gamma
   \log\!\left(P_{ij}^\mathrm{seed}\right)
\end{split}
\end{align}
\noindent where $N$ is the number of seed windows in an event and $N_\mathrm{batch}$ is the batch size. The weight parameters $\lambda_{\mathrm{energy}} = 1$, $\lambda_{\mathrm{position}} = 1$, $\lambda_{\mathrm{seed}} = 0.05$ and the focal loss parameters $\alpha=2$ and $\gamma=0.25$ are adopted from~\cite{dcluster}, since these values were found to provide stable training and satisfactory performance for \mbox{ClusTEX} without requiring a dedicated re-optimisation.

\subsection{Evaluation}
\label{sec:metrics}

A generated-to-reconstructed matching procedure is implemented to associate network predictions with generated photons and to compute performance metrics such as signal efficiency and energy and position resolutions. For each reconstructed object $i$ passing a selection on $P^i_\mathrm{seed}$, a matching criterion $r_\mathrm{match}$ is defined as
\begin{equation}
r^i_\mathrm{match}=
\sqrt{
\frac{(x^i_\mathrm{reco}-x_\mathrm{gen})^2}{\sigma_x^2}
+
\frac{(y^i_\mathrm{reco}-y_\mathrm{gen})^2}{\sigma_y^2}
+
\frac{(E^i_\mathrm{reco}-E_\mathrm{gen})^2}{\sigma_E^2}.
}
\end{equation}

\noindent where $\sigma_x$, $\sigma_y$, and $\sigma_E$ denote the position and energy resolutions of each model. The generated photon that minimises $r^i_\mathrm{match}$ is assigned as the match to the reconstructed object $i$. Cases in which multiple reconstructed objects are assigned to the same generated photon are referred to as \textit{splitting}; it is a consequence of joint training of 1-photon and 2-photon samples.

The signal efficiency is defined as the fraction of matched generated photons relative to the total number of generated photons in the sample. The energy and position resolutions arise from the central 68\% interval of the $X_\mathrm{reco}-X_\mathrm{gen}$ distribution, where $X$ is the quantity of interest. The background rate is quoted per 100,000 events as the number of reconstructed objects that are located more than 3 crystals away from the generated photon.

A dedicated study of computational efficiency and scaling with parameter count is left for future work. As an initial indication of computational performance, the inference times are measured for the final models of the ECAL-inspired topology using a single NVIDIA~A100~GPU with a batch size of~512 in \mbox{TensorFlow 2.6} inference mode. ClusTEX achieves an inference latency of 0.68~ms/event, which is slower but comparable to 0.39 and 0.28~ms/event achieved by SF+GAT and SF+GNN respectively.

\section{Results and discussion}\label{sec:results}
This section compares the performance of the proposed deep learning-based clustering approaches to the PFClustering baseline, first in the toy calorimeter and then in the ECAL-inspired topology. The comparison focuses on (i) signal efficiency, (ii) energy and position resolutions and (iii) the splitting rate, defined as the fraction of generated photons that are matched to multiple reconstructed objects. The quoted resolutions are inclusive of the full $E_\mathrm{gen}$ range and a more detailed breakdown of energy and position resolution as a function of $E_\mathrm{gen}$ is provided in Appendix~\ref{sec:evolution}, where the performance gain of the PoEN models over PFClustering is most pronounced in the two-photon topology at high energies. Unless stated otherwise, all two-step models are evaluated at a common signal-efficiency working point by applying the same SeedFinder requirement and by choosing the PoEN selection such that GAT-PoEN and DW-PoEN are compared at identical efficiency.

\subsection{Photon samples with the toy calorimeter}

The performance comparison between the two-step PoEN models and PFClustering is summarized in Table~\ref{tab:comparison}. Both ML-based approaches significantly outperform PFClustering in the 2-photon topology while preserving the reconstruction quality on the 1-photon sample. In particular, the poor per-photon energy resolution observed for PFClustering in the 2-photon sample reflects the known difficulty of resolving individual photons in overlapping-shower configurations. In contrast, both GNN-based and attention-based PoEN models maintain high per-photon efficiency and substantially improved energy and position resolution in the same regime.

\begin{table}[ht!]
\begin{subtable}{\textwidth}\centering
\begin{tabular}{@{} l ccc @{}}
    \toprule
    & \multicolumn{3}{c}{1-photon sample} \\
    \cmidrule{2-4}
    Metric 
    & SF+GAT & SF+GNN & PFClustering \\
    \midrule
    Efficiency (\%) & 99.7 & 99.7 & 99.8 \\
    $\sigma_E$ (GeV) & 0.53 & 0.54 & 0.61 \\
    $\sigma_x$ (crystal) & 0.02 & 0.02 & 0.04 \\
    Splitting (\%) & 0.05 & 0.56 & -- \\
    $N_\mathrm{background}/100$k events & 31 & 96 & 323k  \\
    \bottomrule
\end{tabular}
\end{subtable}

\hspace*{\fill}\\\hspace*{\fill}%

\begin{subtable}{\textwidth}\centering
\begin{tabular}{@{} l ccc @{}}
    \toprule
    & \multicolumn{3}{c}{2-photon sample} \\
    \cmidrule{2-4}
    Metric 
    & SF+GAT & SF+GNN & PFClustering \\
    \midrule
    Efficiency (\%) & 98.6 & 98.6 & 82.0 \\
    $\sigma_E$ (GeV) & 0.80 & 0.95 & 6.39 \\
    $\sigma_x$ (crystal) & 0.03 & 0.03  & 0.09 \\
    Splitting (\%) & 0.10 & 0.33 & -- \\
    $N_\mathrm{background}/100$k events & 35 & 106 & 323k \\
    \bottomrule
\end{tabular}
\end{subtable}
\caption{Comparison of performance metrics between GAT-$\text{PoEN}_\mathrm{flat}$, GNN-based PoEN and PFClustering for 1-photon and 2-photon testing samples}
\label{tab:comparison}
\end{table}

\begin{figure}[ht!]
    \centering
    \includegraphics[width=0.4\linewidth]{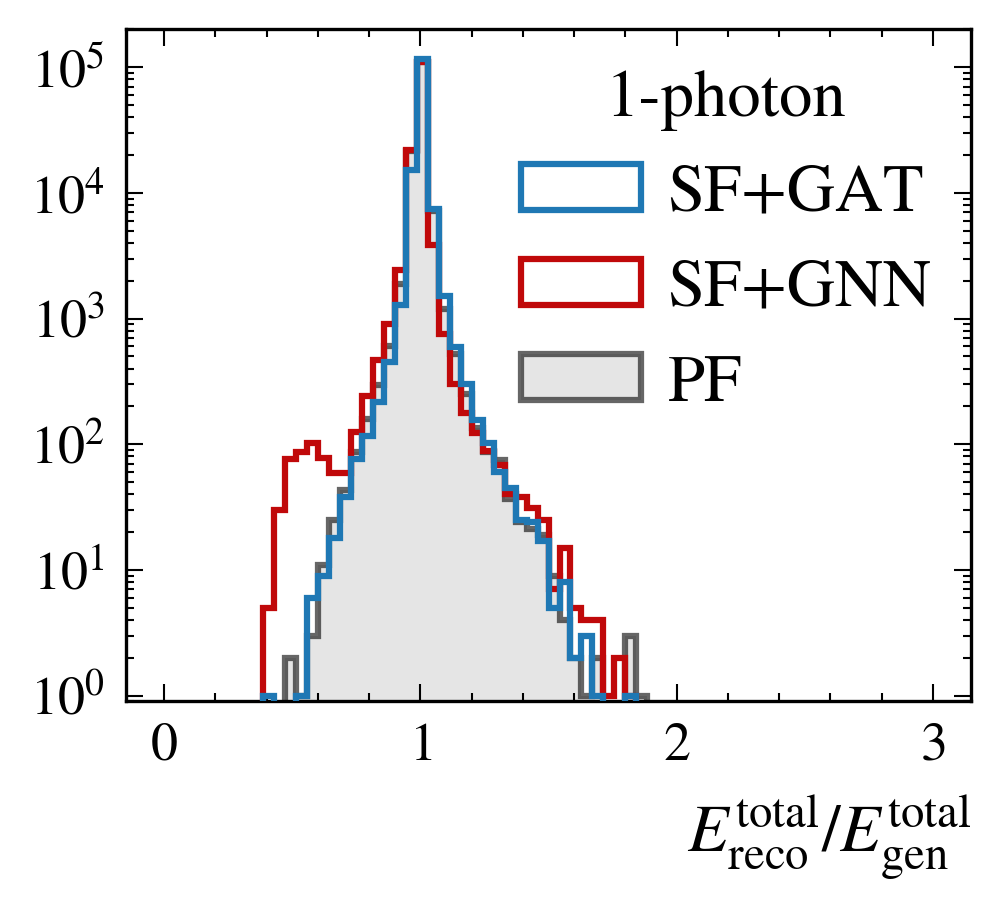}
    \caption{The ratio of the total reconstructed energy to the generated energy for the 1-photon sample as obtained by GAT-$\text{PoEN}_\mathrm{flat}$, GNN-based PoEN and PFClustering~(PF)}
    \label{fig:ratio_split}
\end{figure}

A key qualitative improvement brought by the attention mechanism is the reduction of splitting and the removal of the need for the ad hoc double-pass mitigation used with distance-driven message passing in \cite{dcluster}. In the 1-photon sample, the GNN-based PoEN exhibits a non-negligible splitting fraction (0.56\%), which manifests as a secondary structure in the distribution of the total reconstructed-to-generated energy ratio, $E_\mathrm{reco}^{\mathrm{total}}/E^\mathrm{total}_\mathrm{gen}$ (Fig.~\ref{fig:ratio_split}). This feature is absent for the attention-based PoEN, indicating that attention-driven information exchange suppresses the formation of spurious additional reconstructed objects. In the 2-photon sample, the same trend is observed, with GAT-PoEN achieving both a lower splitting fraction and a better energy resolution than the GNN-based PoEN at the matched-efficiency operating point (Table~\ref{tab:comparison}).
Overall, among the tested two-step approaches, the attention-based PoEN provides the best combined performance, improving the energy resolution for overlapping-shower topologies and substantially mitigating splitting without degrading the reconstruction in the isolated-photon case.

\subsection{Pion sample with the toy calorimeter}
\label{subsec:toy_pion}

To probe the ability to resolve boosted di-photon decays, a neutral pion sample is generated by simulating \pizgg decays incident on the toy calorimeter. A total of 180,000 events with a uniform $\pi^0$ energy distribution in the range 1 to 100~GeV is produced, excluding cases where both photons enter the calorimeter in the same crystal. This sample is used for evaluation only.

\begin{figure}[ht!]
    \centering
    \includegraphics[width=0.6\linewidth]{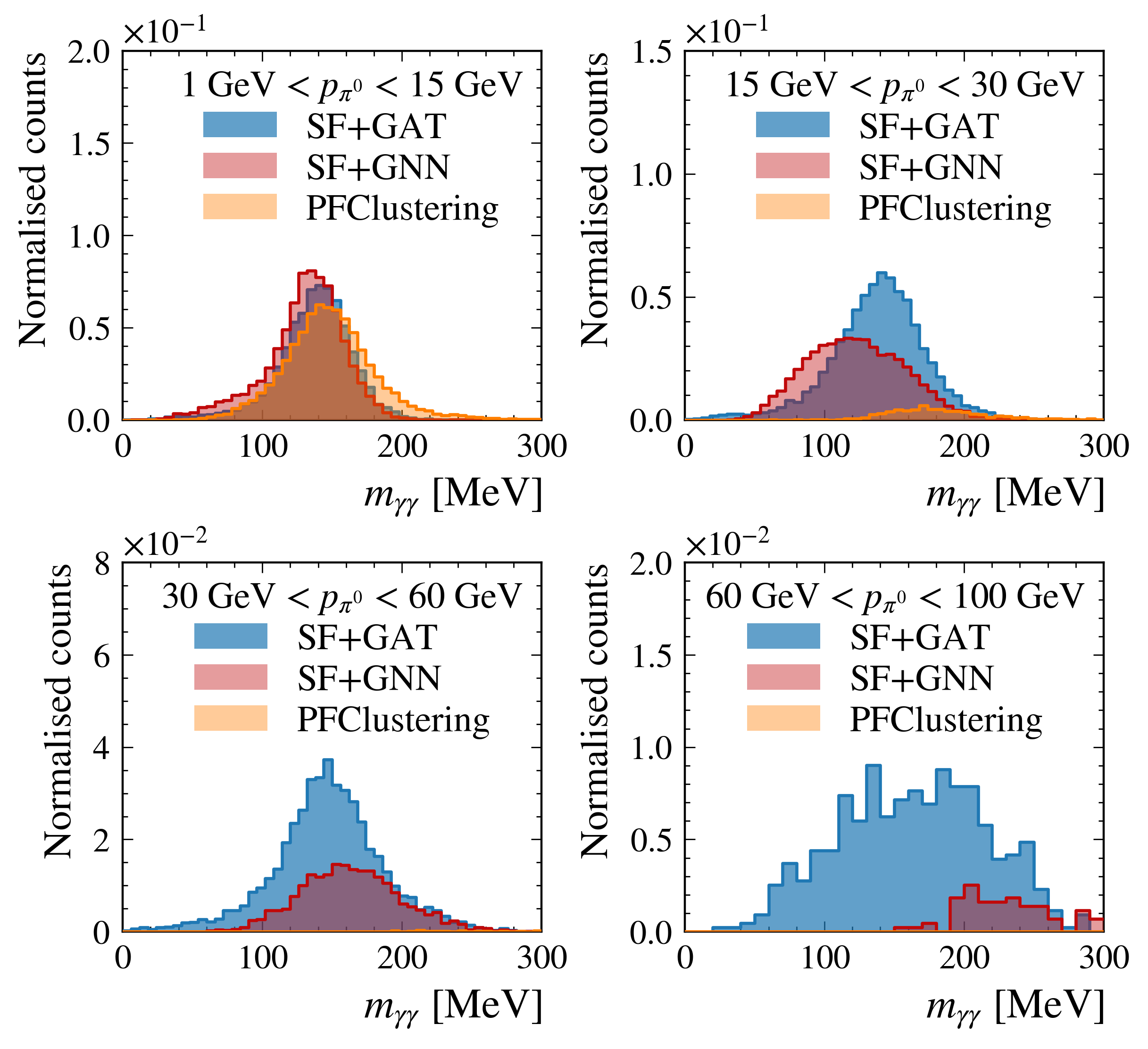}
    \caption{Di-photon invariant mass reconstructed from the $\pi^0$ sample and divided into four regions according to pion momentum. Comparison between GAT-$\text{PoEN}_\mathrm{flat}$, GNN-based PoEN and PFClustering}
    \label{fig:mgg}
\end{figure}

The invariant mass of reconstructed photons is computed as
\begin{equation}
m_{\gamma\gamma}=\sqrt{2E^1_\mathrm{reco}E^2_\mathrm{reco}\left(1-\cos\theta_{12}\right)},
\end{equation}
where $E^1_\mathrm{reco}$ and $E^2_\mathrm{reco}$ are the reconstructed photon energies and $\theta_{12}$ is the opening angle inferred from the reconstructed positions. Figure~\ref{fig:mgg} shows the reconstructed di-photon mass in four pion-momentum regimes. In the boosted regime ($p_{\pi^0}>30$~GeV), only the ML-based approaches retain $\pi^0$ reconstruction capability, while PFClustering becomes inefficient due to the increasing overlap between the two showers. In this regime, the attention-based PoEN is the best performing among the tested methods both in terms of efficiency and mass resolution reconstruction.

\FloatBarrier

\subsection{Photon samples with the ECAL-inspired calorimeter}

\label{subsec:ecal_photon}

The ECAL-inspired simulation introduces reconstruction challenges that are absent in the toy setup. These include more localized shower cores and an increased seed-assignment ambiguity due to the explicit $\eta$ dependence induced by the quasi-projective geometry and the $(\phi,z)$ crystal tilt. In particular, a larger fraction of events exhibit a mismatch between the photon impact point and the crystal with the maximum deposited energy. In this configuration, the convolutional flavour of PoEN has a tendency to outperform GAT-PoEN$_\mathrm{flat}$ due to its local receptive field and translational invariance across the seed window. The single-step transformer (\mbox{ClusTEX}) provides the most robust performance, motivated by (i) its geometrical awareness for the micro and macro features through novel positional encoding and (ii) the removal of cut-driven inefficiencies associated with selecting a fixed number of candidate windows in a two-step pipeline.

\begin{figure}[ht!]
    \centering
    \includegraphics[width=0.8\linewidth]{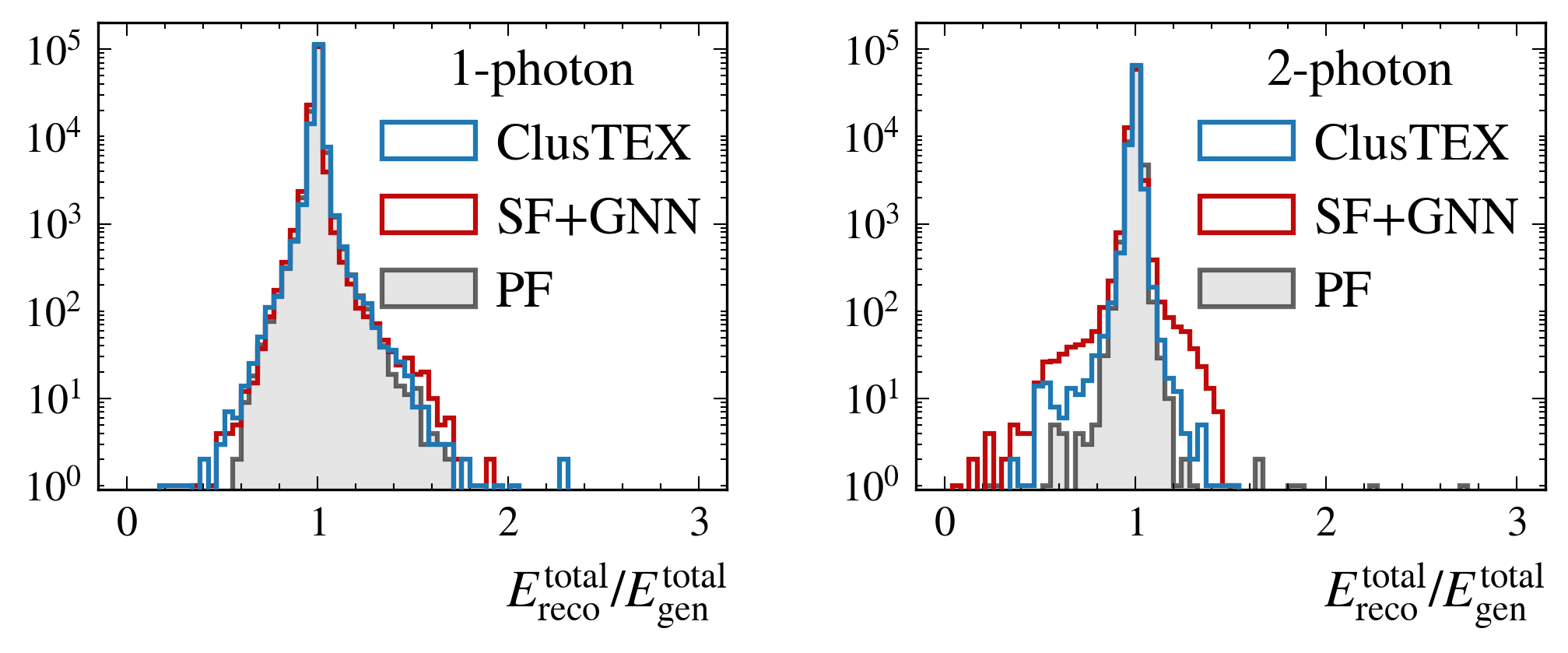}
    \caption{ECAL-inspired topology: ratio of total reconstructed energy to generated energy, $E_\mathrm{reco}^{\mathrm{total}}/E_\mathrm{gen}$, for the 1-photon (left) and 2-photon (right) samples. Comparison between \mbox{ClusTEX}, GNN-based PoEN and PFClustering~(PF)}
    \label{fig:ecal_ratio}
\end{figure}

\begin{figure*}[ht!]
    \centering
    \begin{subfigure}[t]{0.45\linewidth}
        \centering
        \includegraphics[width=0.95\linewidth]{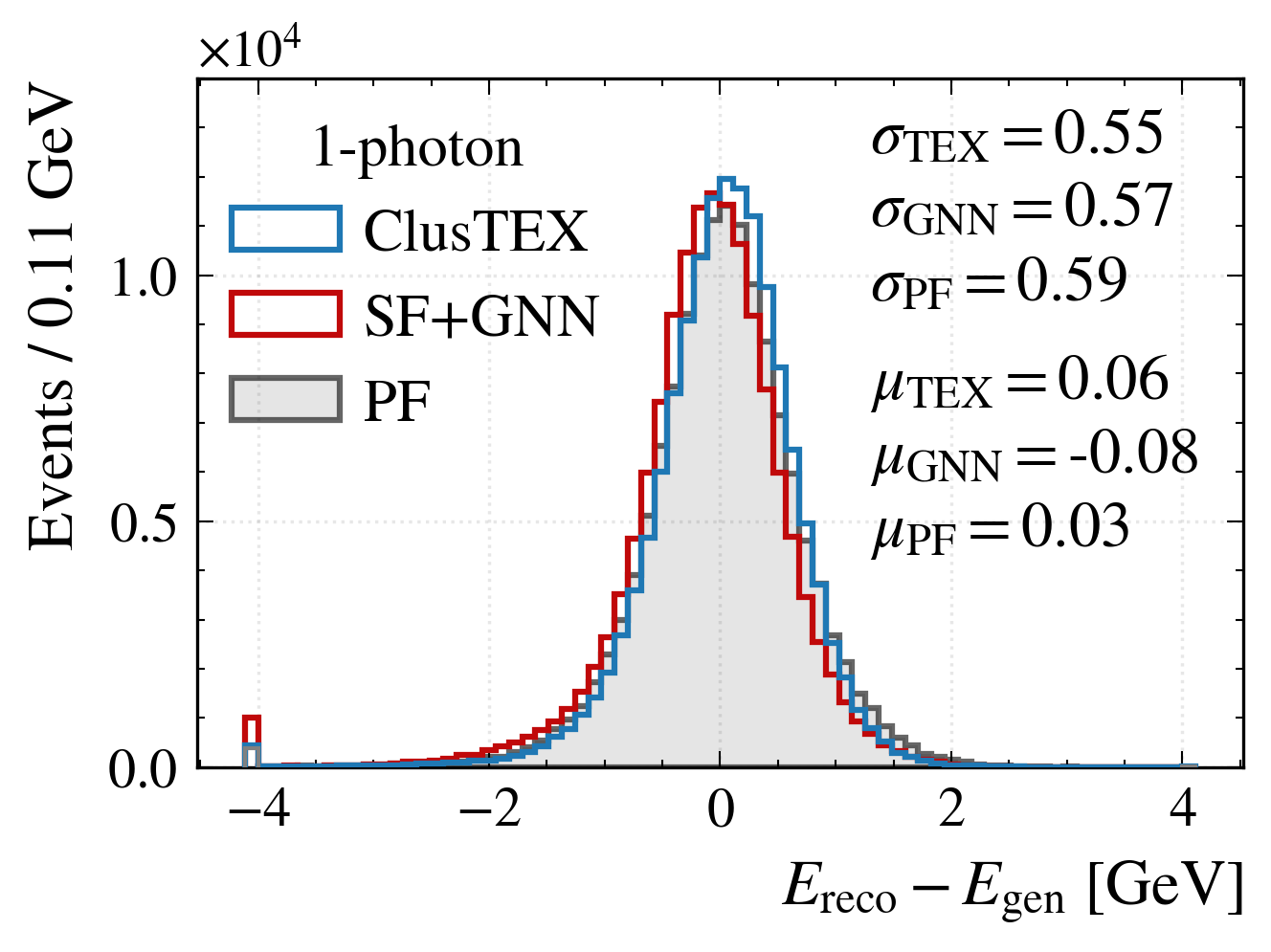}
    \end{subfigure}%
    ~
    \begin{subfigure}[t]{0.45\linewidth}
        \centering
        \includegraphics[width=0.95\linewidth]{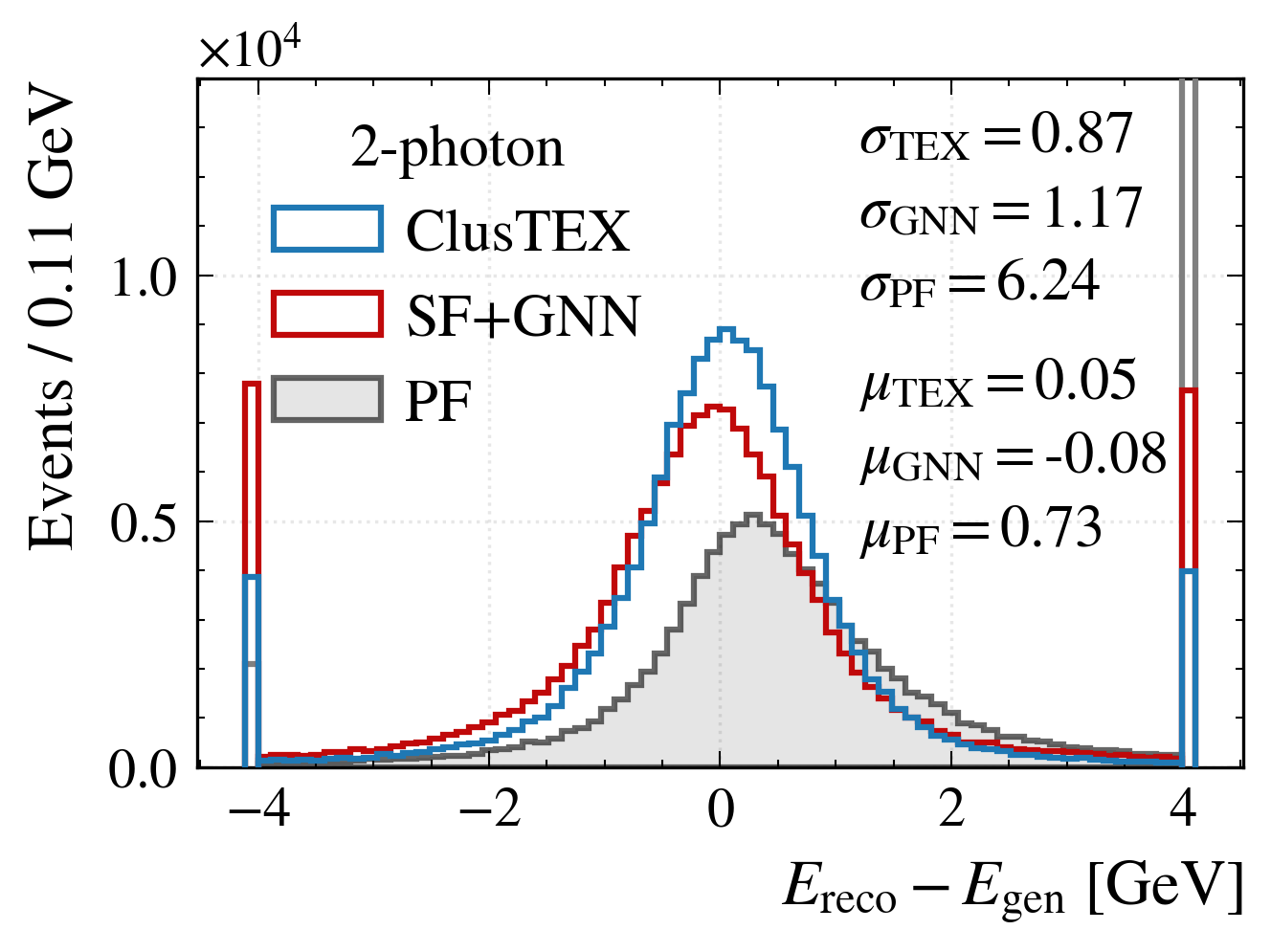}
    \end{subfigure}
    \caption{ECAL-inspired topology: per-photon energy residuals for the 1-photon (left) and 2-photon (right) samples. Comparison between \mbox{ClusTEX}, GNN-based PoEN and PFClustering~(PF)}
    \label{fig:ecal_energy_diff}
\end{figure*}

In order to provide a more complete overview, Table~\ref{tab:comparison_2025} compares \mbox{ClusTEX} to the two-step approaches (GAT-PoEN$_\mathrm{conv}$ and GNN-based PoEN) and to PFClustering. As in the toy calorimeter, the ML-based models significantly outperform PFClustering for overlapping-shower topologies. Relative to the GNN-based PoEN, \mbox{ClusTEX} achieves a markedly improved 2-photon energy resolution and strongly suppresses both the splitting and the background. This is further illustrated by the reconstructed-to-generated energy ratio distributions in Fig.~\ref{fig:ecal_ratio}, where the \mbox{ClusTEX} distribution remains more concentrated around unity than the GNN-based PoEN in both the 1-photon and 2-photon samples. The per-photon energy residuals shown in Fig.~\ref{fig:ecal_energy_diff} similarly demonstrate the improvement in energy resolution, particularly for the overlapping-shower case, relative to both GNN-based PoEN and PFClustering.

\begin{table}[ht!]
\begin{subtable}{\textwidth}\centering
\begin{tabular}{@{} l cccc cccc @{}}
    \toprule
    & \multicolumn{4}{c}{1-photon sample} \\
    \cmidrule{2-5} \cmidrule{6-9}
    Metric
    & \mbox{ClusTEX} & SF+GAT & SF+GNN & PFClustering \\
    \midrule
    Efficiency (\%) & 99.7 & 99.8 & 99.8 & 99.8 \\
    $\sigma_E$ (GeV) & 0.55 & 0.58 & 0.57 & 0.59 \\
    $\sigma_{i\phi}$ (crystal) & 0.03 & 0.03 & 0.03 & 0.05 \\
    $\sigma_{i\eta}$ (crystal) & 0.04 & 0.04 & 0.04 & 0.06 \\
    Splitting (\%) & 0.08 & 0.17 & 0.30 & -- \\
    $N_\mathrm{bkg}/100$k events & 73 & 26 & 102 & 312k \\
    \bottomrule
\end{tabular}
\end{subtable}

\hspace*{\fill}\\\hspace*{\fill}%

\begin{subtable}{\textwidth}\centering
\begin{tabular}{@{} l cccc cccc @{}}
    \toprule
    & \multicolumn{4}{c}{2-photon sample} \\
    \cmidrule{2-5} \cmidrule{6-9}
    Metric
    & \mbox{ClusTEX} & SF+GAT & SF+GNN & PFClustering \\
    \midrule
    Efficiency (\%) & 98.7 & 98.3 & 98.3 & 82.0 \\
    $\sigma_E$ (GeV) & 0.87 & 1.10 & 1.17 & 6.24 \\
    $\sigma_{i\phi}$ (crystal) & 0.04 & 0.04 & 0.04 & 0.09 \\
    $\sigma_{i\eta}$ (crystal) & 0.04 & 0.04 & 0.04 & 0.09 \\
    Splitting (\%) & 0.06 & 0.31 & 0.59 & -- \\
    $N_\mathrm{bkg}/100$k events & 99 & 84 & 178 & 311k \\
    \bottomrule
\end{tabular}
\end{subtable}
\caption{ECAL-inspired topology: comparison of performance metrics between the single-step transformer (\mbox{ClusTEX}), GAT-PoEN$_\mathrm{conv}$, GNN-based PoEN and PFClustering for the 1-photon and 2-photon testing samples}
\label{tab:comparison_2025}
\end{table}

\FloatBarrier

\subsection{Pion sample with the ECAL-inspired calorimeter}
\label{subsec:ecal_pion}

\begin{figure}[ht!]
    \centering
    \includegraphics[width=0.7\linewidth]{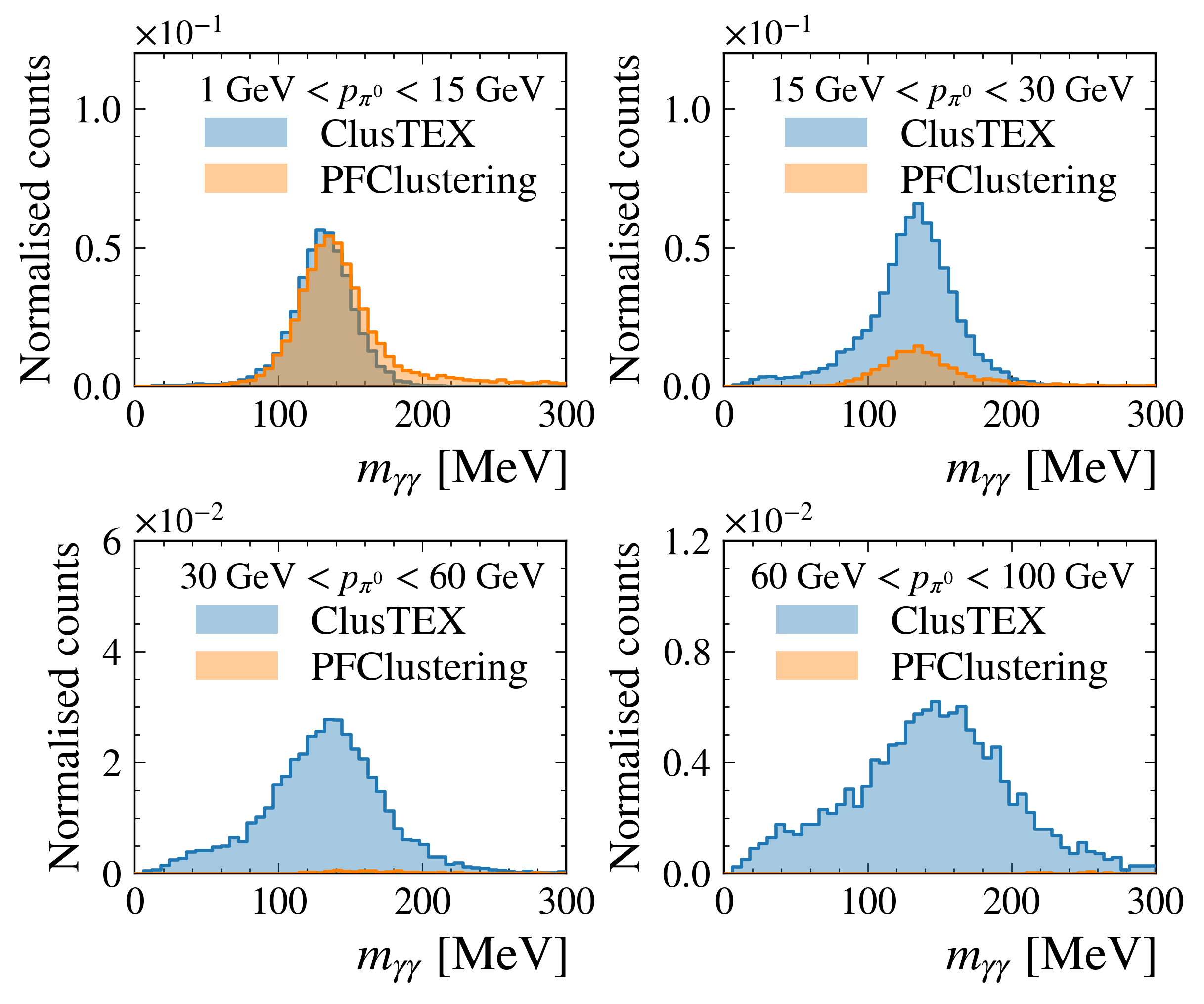}
    \caption{ECAL-inspired topology: reconstructed di-photon invariant mass from the $\pi^0$ sample, divided into four regions according to pion momentum}
    \label{fig:mgg_ecal}
\end{figure}

The neutral-pion evaluation is repeated in the ECAL-inspired topology to assess the robustness of boosted di-photon reconstruction under geometry-driven nonuniformities and seed ambiguities. The reconstructed di-photon invariant mass distributions are shown in Fig.~\ref{fig:mgg_ecal}. The \mbox{ClusTEX} model retains stable reconstruction performance across the pion-momentum regimes, in line with its improved handling of overlaps and its reduced splitting compared to the two-step approaches. Having a handle on these decays is also important for $\tau$-lepton reconstruction, as \pizgg appears frequently in $\tau$ decay chains. Improved resolution of close-by photons can therefore translate into more accurate reconstruction of hadronic-$\tau$ decay and as a result improved sensitivity in measurements that rely on $\tau$ decay kinematics, including polarization-related observables. These results also indicate strong potential for heavier neutral mesons decaying to photon pairs, such as $\eta\!\to\!\gamma\gamma$, where the larger particle mass leads to a larger average opening angle at a given transverse momentum; therefore, a higher expected reconstruction efficiency. This is particularly interesting for heavy-ion analyses, where reliable reconstruction of neutral resonances in high-occupancy environments is a central experimental challenge. 

\FloatBarrier

\subsection{Compensation of the non-responsive readout regions}
\label{subsec:dead_cells}

\begin{figure*}[ht!]
    \centering
    \begin{subfigure}[t]{0.45\linewidth}
        \centering
        \includegraphics[width=0.95\linewidth]{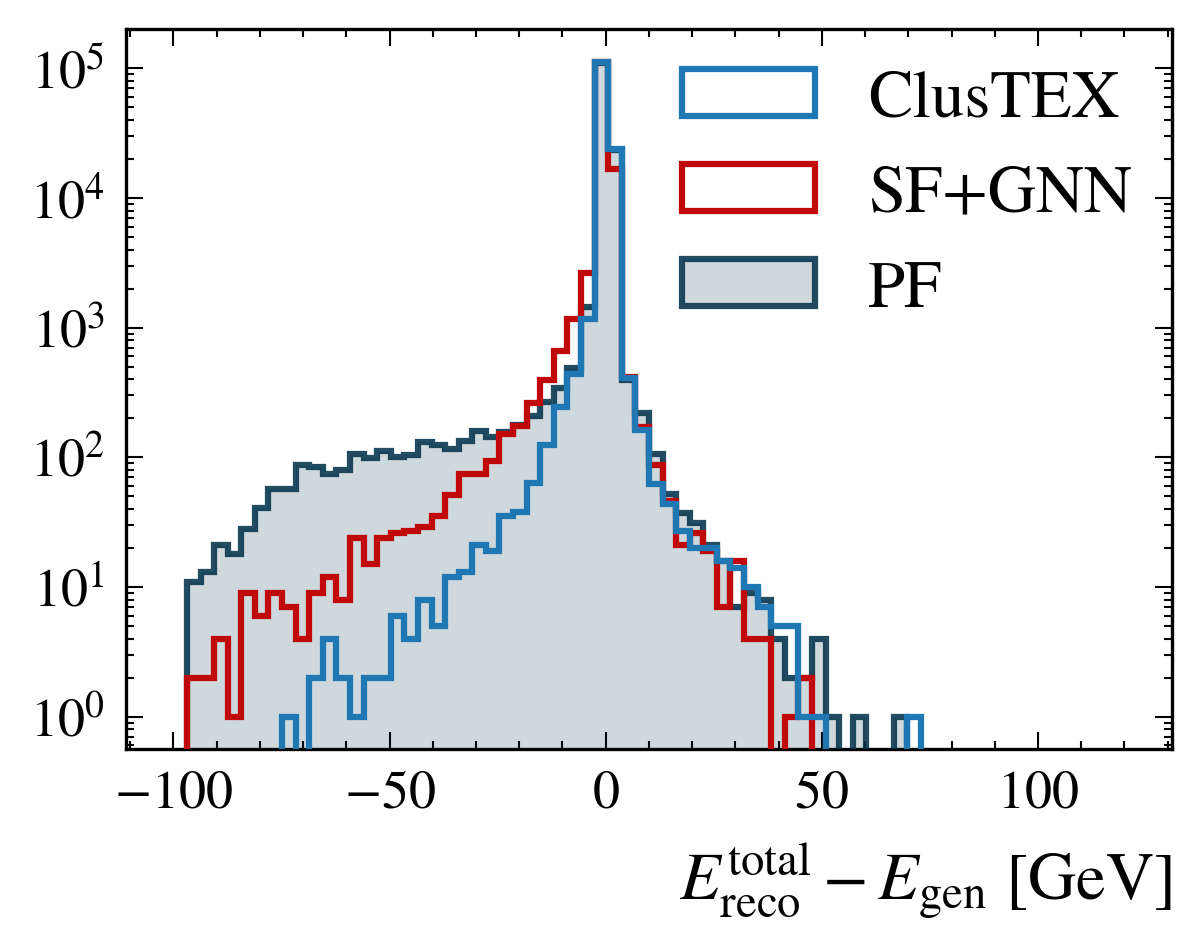}
    \end{subfigure}
    \begin{subfigure}[t]{0.45\linewidth}
        \centering
        \includegraphics[width=0.87\linewidth]{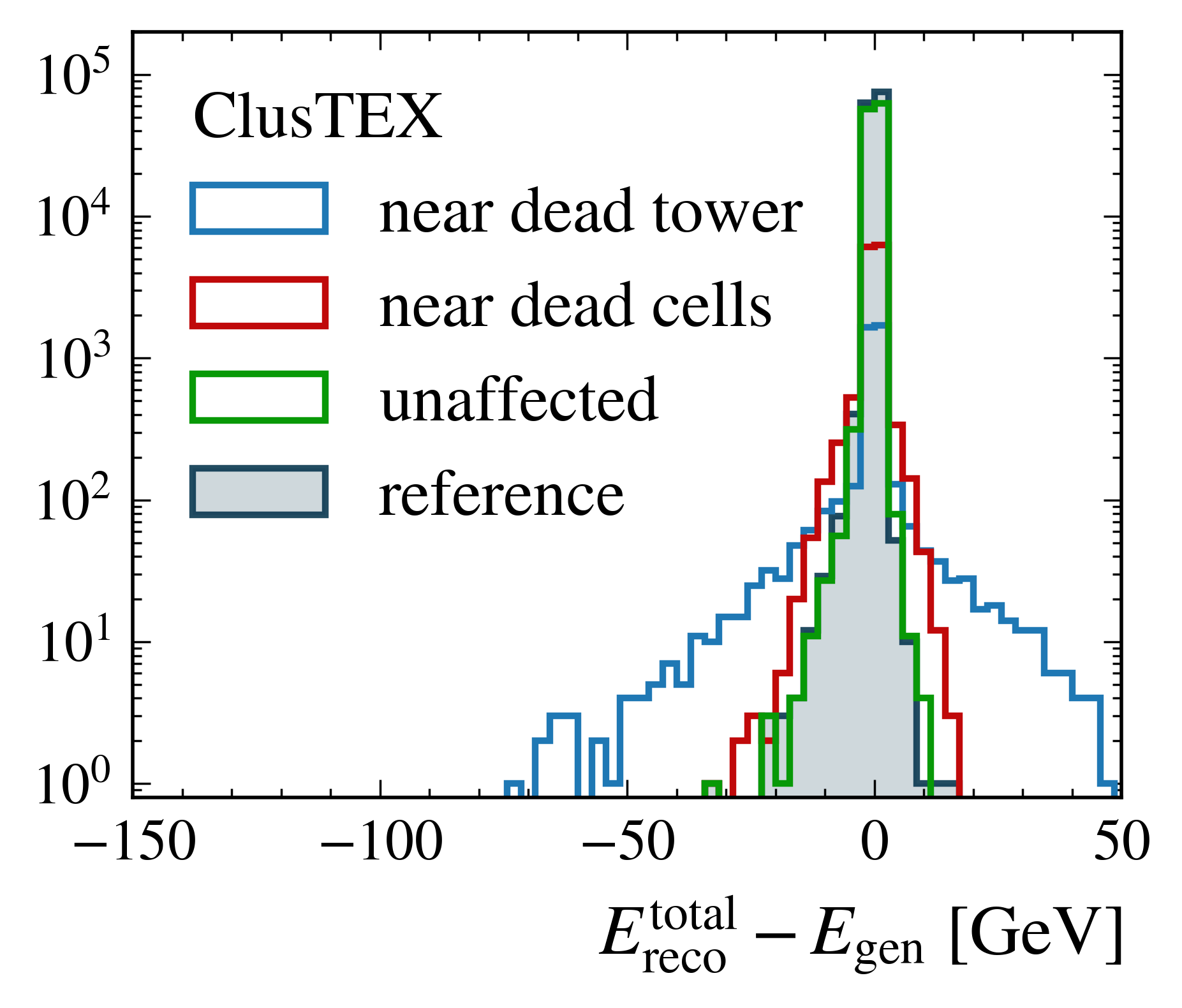}
    \end{subfigure}
    \caption{Impact of disabling 1\% of crystals and one $5\times5$ readout tower on the difference $E_\mathrm{reco}^{\mathrm{total}}-E_\mathrm{gen}$. Left: comparison of \mbox{ClusTEX}, GNN-based PoEN and PFClustering~(PF). Right: \mbox{ClusTEX} reconstruction accuracy based on the photon impact position. Comparison of energy residuals near non-responsive regions, in an unaffected region far from the dead crystals and the nominal result}
    \label{fig:dead_cells}
\end{figure*}

To study robustness to channel failures and non-responsive readout regions, we simulate detector conditions in which 1\% of crystals are disabled together with one non-responsive $5\times5$ readout tower as discussed in~\ref{sec:dead_channels_posenc}. The corresponding dead channels are present during both the training and evaluation stages for all model comparisons, rather than being introduced only at inference time. Figure~\ref{fig:dead_cells} summarizes the impact on energy reconstruction. While PFClustering and the two-step models exhibit larger distortions of the reconstructed-to-generated energy ratio and broader energy residual distributions, \mbox{ClusTEX} remains more resilient. In particular, the \mbox{ClusTEX} distributions indicate that the single-step transformer can partially compensate for missing energy in non-responsive regions by exploiting contextual information from neighbouring active crystals and from its global detector-position awareness, resulting in a reduced bias and improved stability in the presence of localized failures.

\begin{figure*}[ht!]
    \centering
    \includegraphics[width=0.9\linewidth]{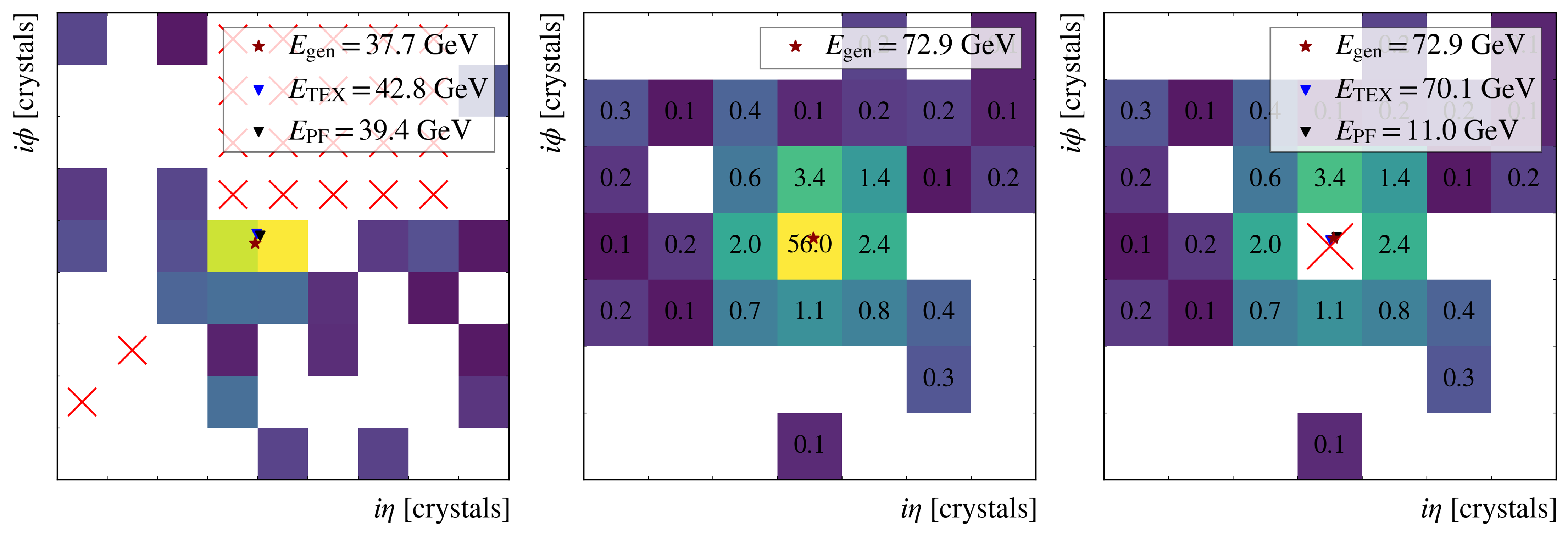}
    \caption{Event displays of seed windows with a significant proportion formed by non-responsive channels}
    \label{fig:dead_event_displays}
\end{figure*}

Event-display examples illustrating seed windows with a significant proportion of dead channels are shown in Fig.~\ref{fig:dead_event_displays}.

\FloatBarrier

\section{Conclusion}
\label{sec:conclusion}

In this paper, we have presented a set of deep learning-based approaches for clustering electromagnetic calorimeter energy deposits and reconstructing the energy and impact position of photons directly from calorimeter readout. The study covers two-step strategies, in which candidate seed windows are first identified and then jointly processed by a reconstruction network, as well as a single-step graph transformer, \mbox{ClusTEX}, that performs candidate selection and kinematic reconstruction in one inference stage once a local graph is formed.

In the toy-calorimeter studies, both message-passing and attention-based two-step models significantly improve reconstruction performance in overlapping-shower topologies relative to the PFClustering baseline while preserving the reconstruction quality for isolated photons. Replacing distance-driven aggregation by attention reduces spurious reconstructions and suppresses splitting, removing the need for ad hoc multi-pass inference and yielding improved energy reconstruction in the 2-photon topology. 

In the ECAL-inspired topology, where geometry-induced nonuniformities and an explicit $(\eta,\phi)$ dependence introduce additional seed ambiguities and position-dependent effects, the single-step transformer provides the most robust overall performance with the highest achieved precision. \mbox{ClusTEX} yields improved energy resolution and a strong reduction of splitting compared to two-step approaches and PFClustering. Furthermore, it maintains high reconstruction efficiency in close-by photon event topologies. These improvements translate into boosted di-photon signatures: the ML-based approaches retain \pizgg reconstruction capability in regimes where PFClustering becomes inefficient due to increasing shower overlap, with the attention-based model providing the most stable performance. This mechanism is especially relevant for heavy-flavour measurements in extremely high-occupancy environments, for instance study of $\mathrm{b\!\to\!s\gamma}$ transitions, where close-by $\gamma$, $\pi^0$, $\eta$ occur frequently in b-initiated jets.\\
Lastly, the dedicated studies with non-responsive readout regions further show that \mbox{ClusTEX} remains stable under localized detector failures and can partially compensate for the missing energy by exploiting contextual information and detector-position awareness enabled by its novel positional encoding approach.

The results demonstrate that graph transformer-based clustering with an explicit separation of local and global positional information is a promising direction for electromagnetic reconstruction in high-occupancy environments. By improving the reconstruction of overlapping electromagnetic showers and providing robustness to detector nonuniformities and channel failures, the proposed approach offers a scalable foundation for future developments toward HL-LHC reconstruction conditions and for downstream steps such as superclustering.

\subsection*{Acknowledgements}

This research work is supported by the European Union's Horizon Europe research and innovation programme under the Marie Sk\l{}odowska-Curie COFUND grant agreement No 101127936 (DeMythif.AI) and France 2030 funding, managed by the National Research Agency (ANR), as part of IA CLUSTER program, reference ANR-23-IACL-0003 - DATAIA CLUSTER.

\subsection*{Data availability}

The datasets generated for the ECAL-inspired topology are available at \cite{zenodo}.

% Continue figure numbering in appendix
\newcounter{mainfig}
\setcounter{mainfig}{\value{figure}}

\newpage

\begin{appendices}

\renewcommand{\thefigure}{\arabic{figure}}
\setcounter{figure}{\value{mainfig}}

\section{Calorimeter simulations}\label{secA1}

This section summarizes the key differences between the toy calorimeter and the ECAL-inspired simulation. These differences arise from (i) the inclusion of tracker material, (ii) seed assignment ambiguities, (iii) shower asymmetry, and (iv) differences in shower size.

\begin{figure}[ht!]
    \centering
    \includegraphics[width=0.5\linewidth]{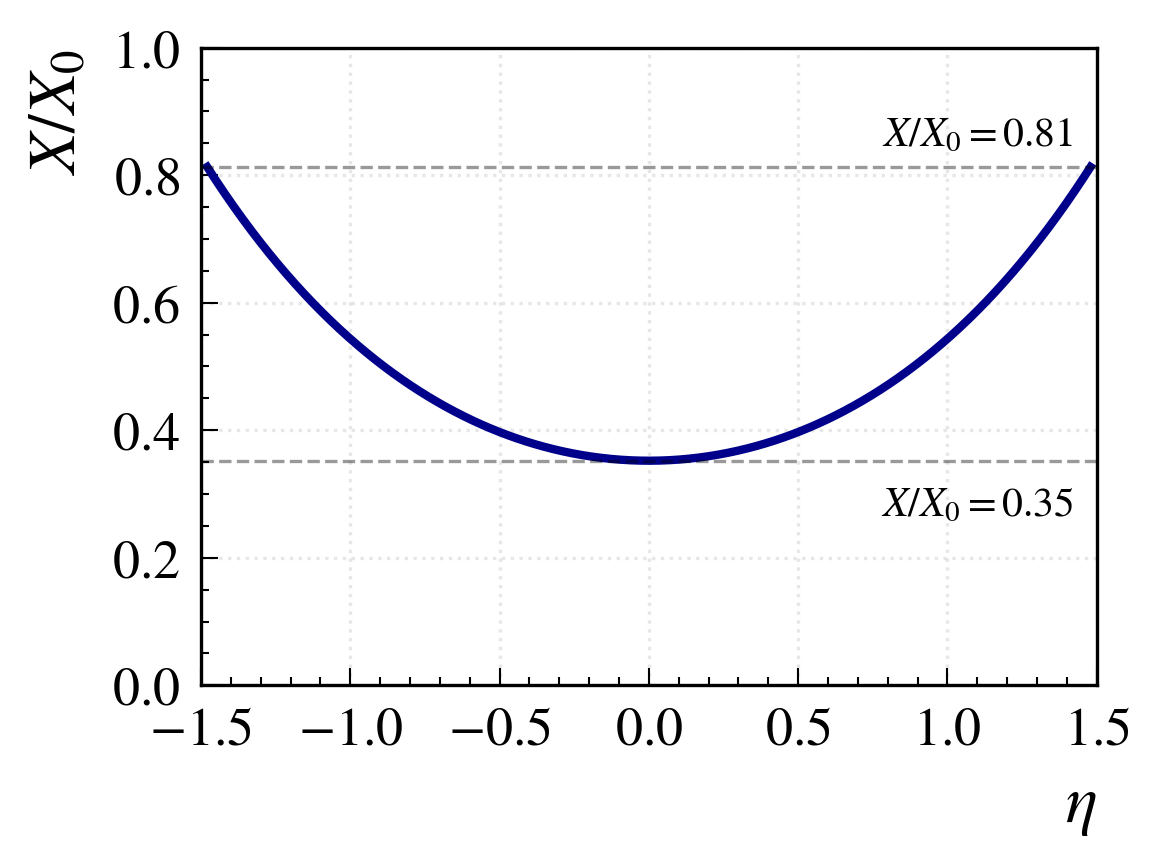}
    \caption{Profile of tracker material against pseudorapidity in units of silicon radiation lengths $X_0=9.370$~cm}
    \label{fig:tracker_material}
\end{figure}

The ECAL-inspired simulation reproduces the detailed crystal geometry of the CMS ECAL barrel but does not explicitly simulate the full tracker system, read-out electronics, or magnetic field. To emulate the material budget upstream of the calorimeter, an equivalent thickness of 33~mm of silicon is placed immediately in front of the scintillating crystal volume.
The resulting material profile as a function of pseudorapidity is shown in Fig.~\ref{fig:tracker_material}, expressed in units of silicon radiation length ($X_0 = 9.370$~cm). The material budget varies with $\eta$ and induces early shower development, thereby modifying both longitudinal and lateral shower profiles compared to the toy calorimeter, in which there is no upstream material.

\begin{figure}[ht!]
    \centering
    \includegraphics[width=0.4\linewidth]{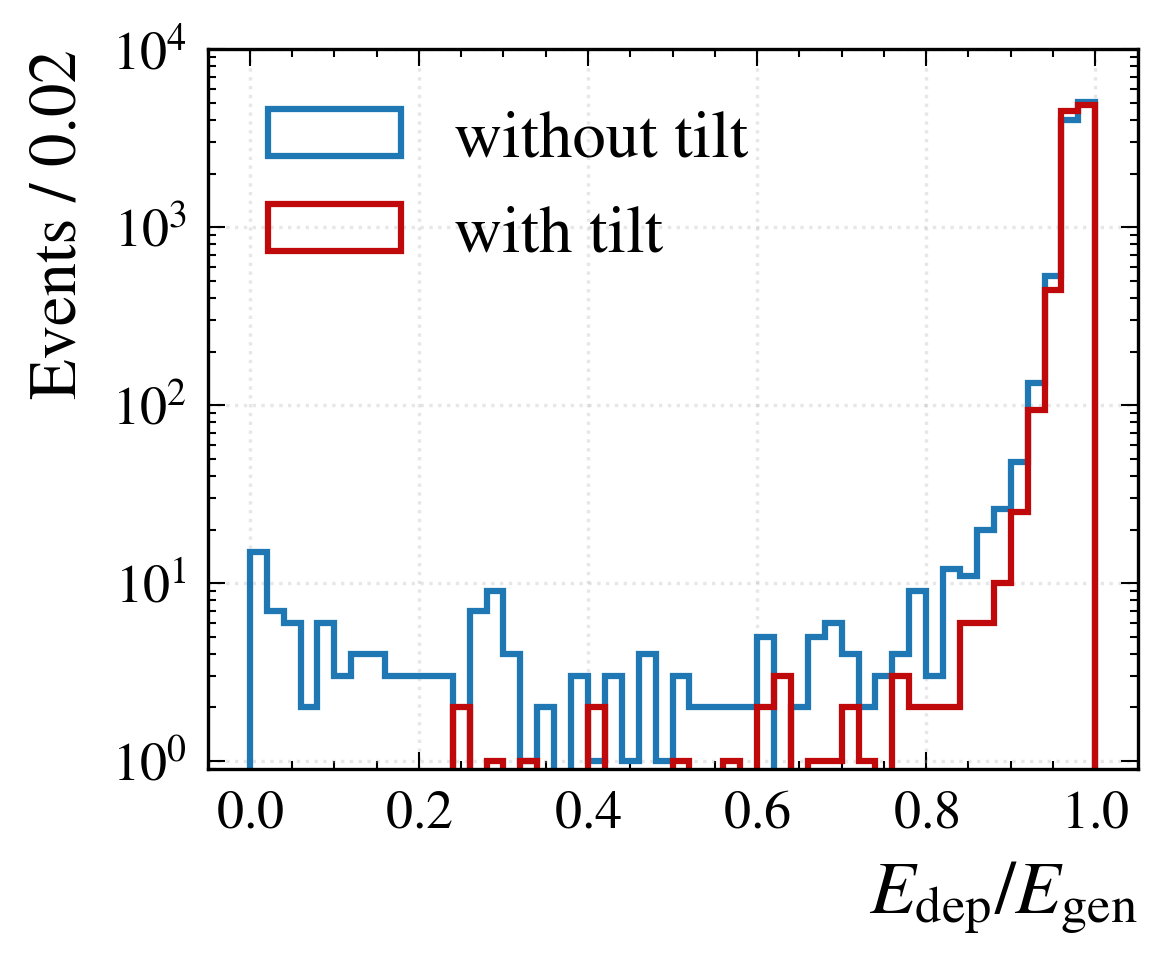}
    \caption{The impact of adding a 3$^\circ$-tilt in $z$ and $\phi$ on the total deposited energy. Based on a simulation of 10,000 single photons in [1,100] GeV range}
    \label{fig:edep_egen_tilt}
\end{figure}

\begin{figure}[ht!]
    \centering
    \includegraphics[width=1\linewidth]{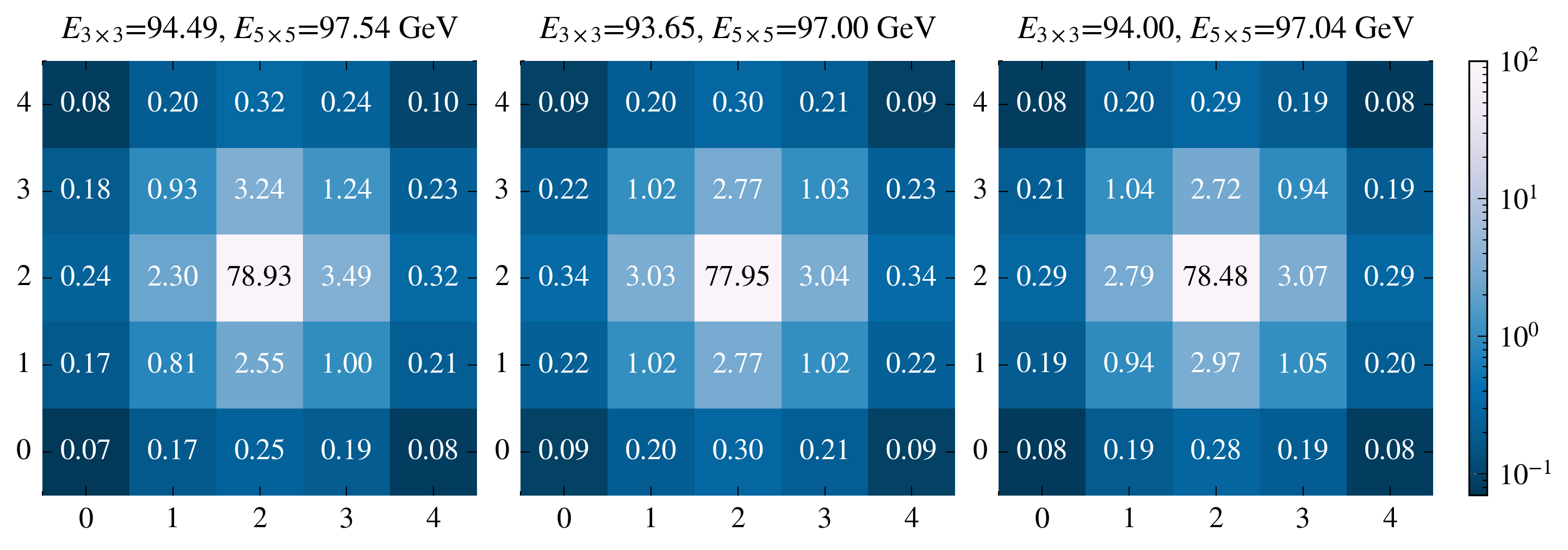}
    \caption{Average energy distribution of 100-GeV electron beam incident on the central crystal of the given 5$\times$5 matrix for ECAL test beam data (left), toy calorimeter (middle) and ECAL-inspired simulation (right). Simulation of 1000 events and test beam data from \cite{tbmatrix}}
    \label{fig:test_beam}
\end{figure}

The crystals are azimuthally and radially tilted towards to the interaction point and have an additional $3^\circ$-tilt in both directions to increase the path length from the interaction point and reduce shower leakage through inter-crystal gaps. The impact of the additional $3^\circ$-tilt on the energy deposited in the crystals is demonstrated in Fig.~\ref{fig:edep_egen_tilt}. This topology also creates an asymmetry in the shower shape as seen in the average energy distribution of 100-GeV electron beam incident on the central crystal of a 5$\times$5 window (Fig.~\ref{fig:test_beam}).

\begin{figure}
    \centering
    \includegraphics[width=0.7\linewidth]{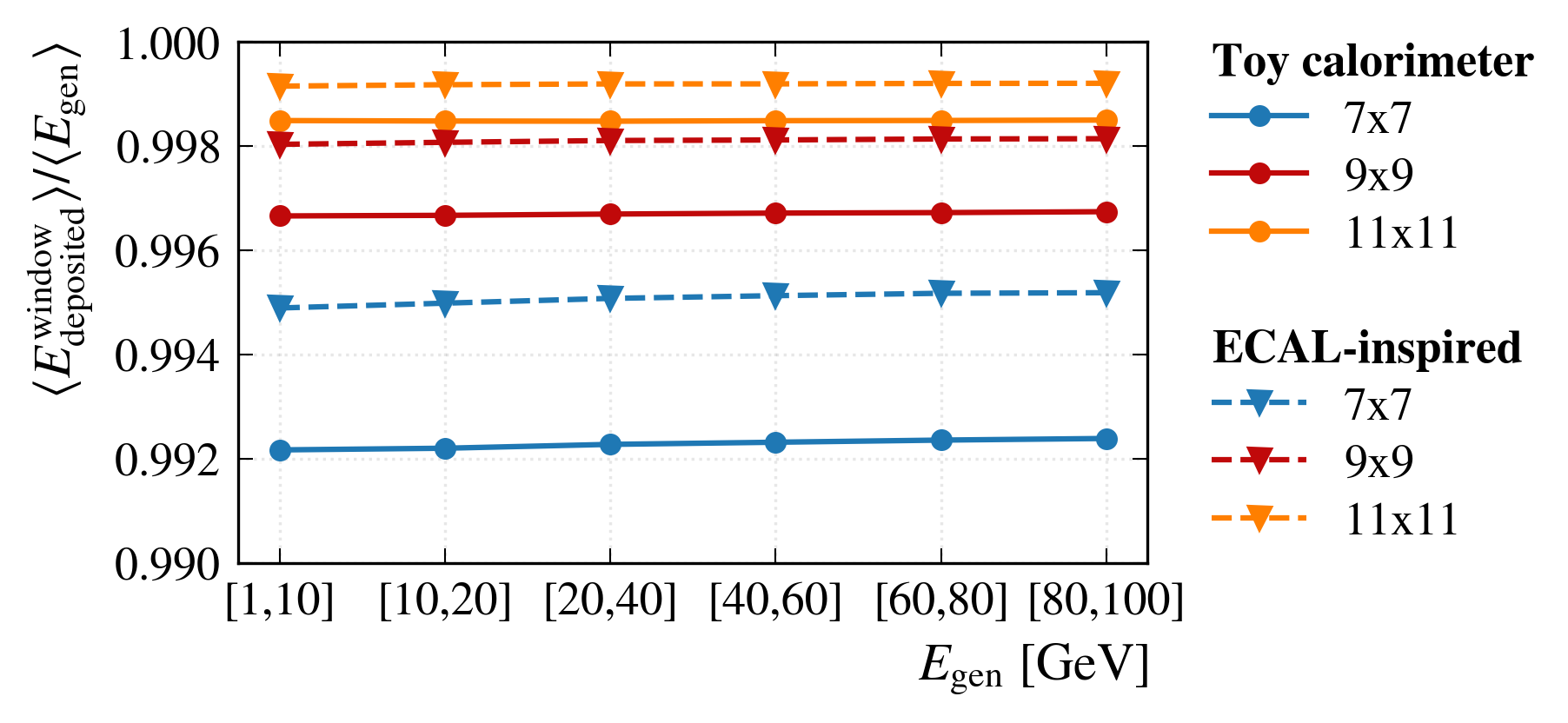}
    \caption{Average fraction of generated energy contained in seed windows of varying sizes for the toy calorimeter and the ECAL-inspired simulation. Based on 100,000 single photons in [1,100] GeV range}
    \label{fig:containment}
\end{figure}

Beyond asymmetry, the ECAL-inspired configuration also produces a more localised shower core compared to the toy calorimeter. As per Fig.~\ref{fig:test_beam}, the energy contained in the central crystal, 3$\times$3 window and 5$\times$5 window is $E=(77.95, 93.65, 97.00)$~GeV for the toy calorimeter and $E=(78.48, 94.00, 97.04)$~GeV for the ECAL-inspired simulation. This difference is consistent with earlier shower development in the upstream material and the modified detector geometry. The same trend is observed for a photon beam across the full energy range and for larger seed windows, as shown in Fig.~\ref{fig:containment}. 

Furthermore, in 13\% of single-photon events generated in the ECAL-inspired topology, the photon impact position does not correspond to the crystal with the highest deposited energy. This fraction is higher compared to the toy calorimeter and this creates a seed assignment ambiguity since typically a seed would coincide with both the impact position and the local maximum.

\FloatBarrier

\section{Non-responsive readout channels and positional encoding}
\label{sec:dead_channels_posenc}

In this section we provide details on two additional studies that clarify two key aspects of the \mbox{ClusTEX} design. First, we study robustness to localized detector non-responses by introducing unresponsive readout channels in the calorimeter window, including a contiguous $5\times5$ region corresponding to one ECAL barrel trigger-tower granularity and additional isolated channels. Such localized failures can arise from detector operations and aging effects, and they represent an important stress test for reconstruction stability under non-ideal conditions. Second, we provide diagnostic plots that illustrate how the global positional embedding layer evolves during training and how different positional encoding strategies affect convergence.

\subsection{Global positional embedding behavior}

The global positional encoding in \mbox{ClusTEX} is implemented as a learnable embedding that is summed to the node representation. While the embedding is learned without any explicit functional constraint, its trained weights can develop smooth, detector-scale structure that reflects the underlying geometry and the coordinate dependence of the shower patterns.

\begin{figure}[ht!]
    \centering
    \includegraphics[width=1\linewidth]{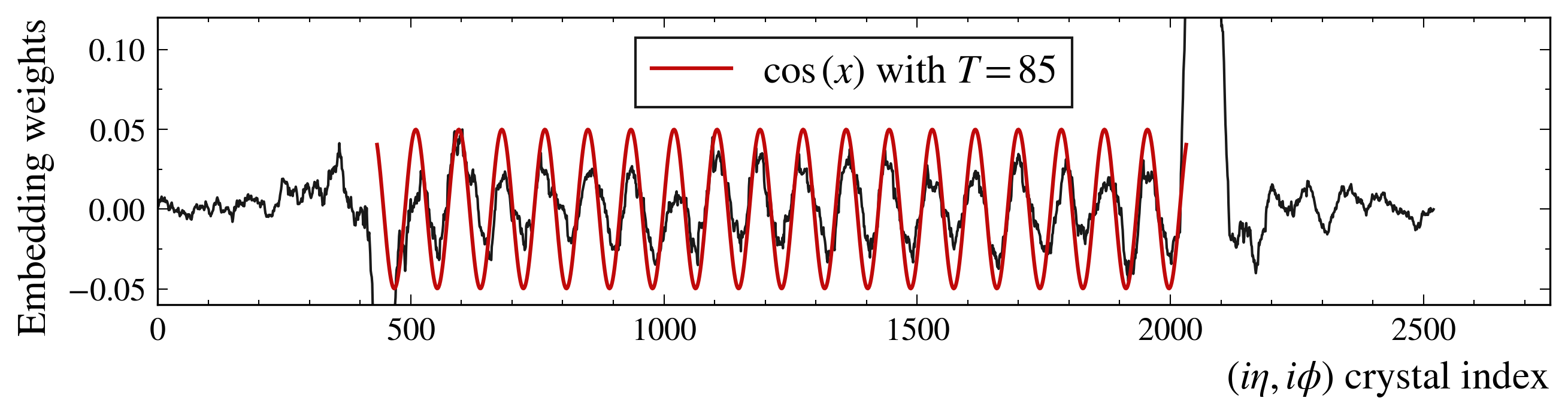}
    \caption{Running average of a single dimension of the global positional embedding-layer weights (black) and a reference cosine function with a period equal to the longitudinal detector extent (red)}
    \label{fig:posenc_weights_cos}
\end{figure}

Figure~\ref{fig:posenc_weights_cos} shows a running average of a single embedding-weight component compared to a reference cosine function with a period equal to the longitudinal extent of the simulated detector segment. The qualitative similarity illustrates that the learned embedding can acquire slowly varying, approximately periodic structure across the detector coordinate, consistent with the role of global positional information in capturing detector-dependent response effects.

\subsection{Impact of positional encoding strategy on convergence}

\begin{figure}[ht!]
    \centering
    \includegraphics[width=0.9\linewidth]{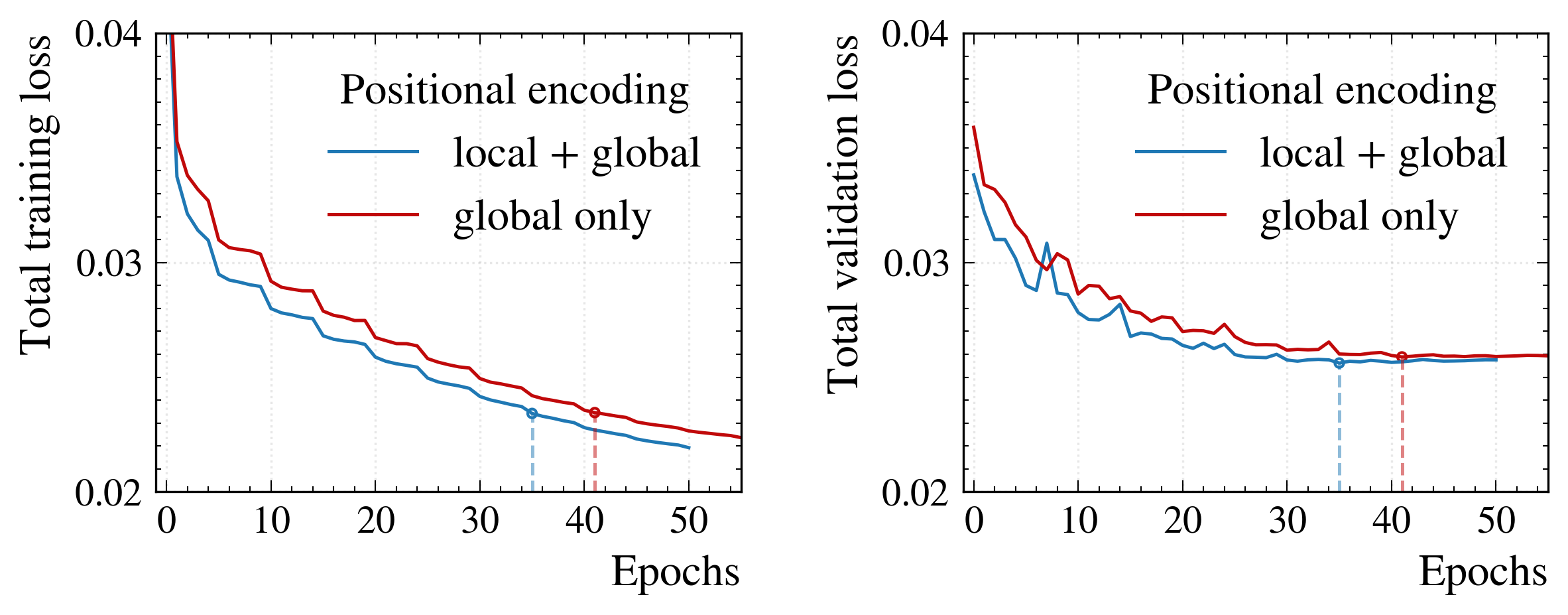}
    \caption{Total training (left) and validation (right) loss for \mbox{ClusTEX} trained with different positional encoding strategies. The local+global strategy converges earlier and reaches the lowest validation loss}
    \label{fig:posenc_loss}
\end{figure}

To quantify the training-level impact of positional encoding, \mbox{ClusTEX} is trained with multiple positional encoding configurations and the resulting training and validation losses are compared. Figure~\ref{fig:posenc_loss} shows that the novel local+global encoding strategy achieves both improved convergence in training and a lower best validation loss than the alternative configuration. This is consistent with the intended functional separation: the local encoding provides an anchor-centered coordinate system inside each overlap window, while the global encoding provides absolute detector awareness through a summed embedding. The combination accelerates optimization and improves the quality of the final model.

\subsection{Non-responsive readout channels configuration}

To study reconstruction robustness to localized detector failures, a subset of channels in the calorimeter window is set to be non-responsive. Concretely, the measured energy in the affected channels is forced to zero before constructing seed windows and graphs. The configuration includes (i) one contiguous $5\times5$ region corresponding to the ECAL barrel trigger-tower granularity and (ii) additional isolated non-responsive channels distributed within the window.

\begin{figure}[ht!]
    \centering
    \includegraphics[width=0.6\linewidth]{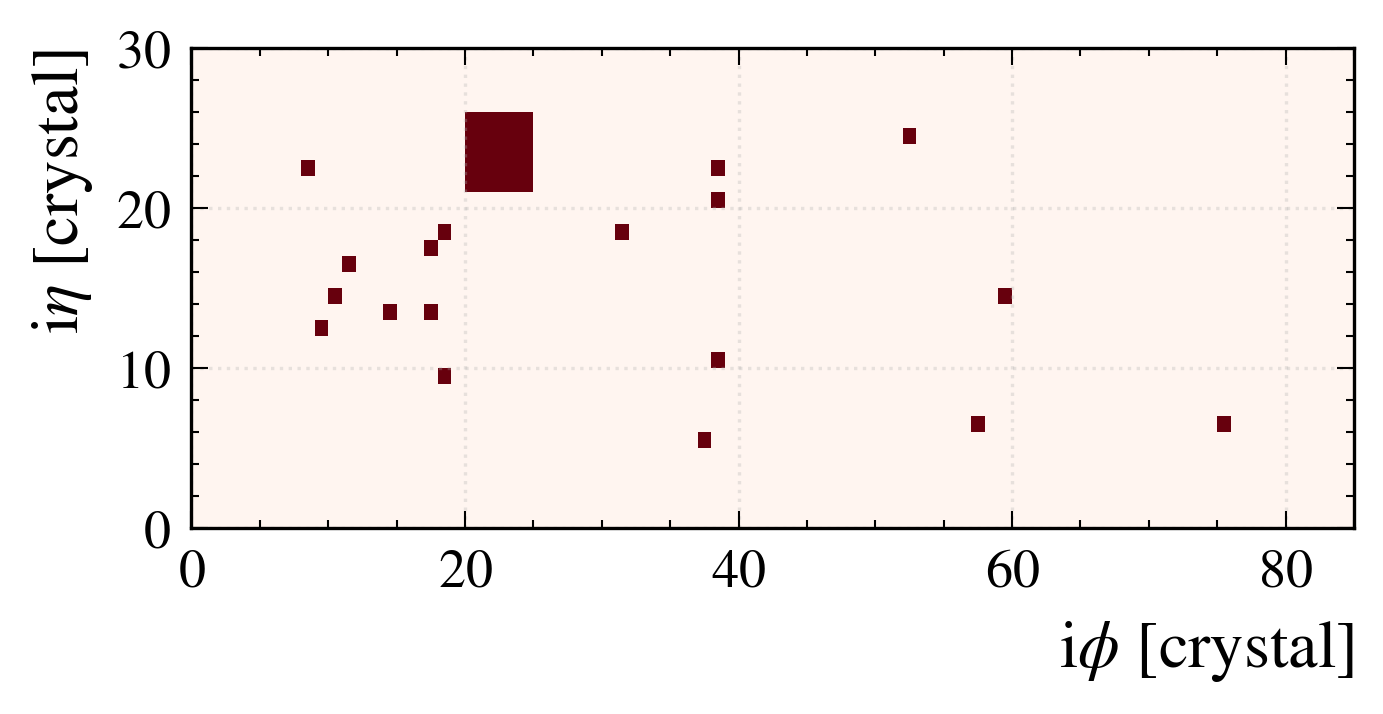}
    \caption{The map showing 18 isolated non-responsive channels and one non-responsive $5\times5$ readout region corresponding to one ECAL barrel trigger-tower granularity, which is used to mimic the realistic operation conditions}
    \label{fig:kill_map}
\end{figure}

Figure~\ref{fig:kill_map} shows an example calorimeter window with 18 isolated non-responsive channels and one non-responsive $5\times5$ readout region. This configuration is used in the non-responsive-channel performance studies presented in Section~\ref{subsec:dead_cells}.

\FloatBarrier

\section{PFClustering algorithm}
\label{sec:pfclusteringalgo}

While PFClustering provides a robust baseline for isolated deposits, its clustering strategy is intrinsically challenged when two electromagnetic showers overlap within the same local region. In such cases, the seed assignment becomes ambiguous and the subsequent energy-sharing procedure can fail to correctly attribute energy deposits to the individual initiating photons. This shows itself as a reduced per-photon reconstruction efficiency at low photon energies and a strong degradation of the per-photon energy resolution compared to the deep learning models, which jointly reason over multiple candidates without requiring a local-maximum constraint.

At a high level, PFClustering proceeds by selecting localized energy deposits above a per-crystal threshold and grouping them into topological structures built around local maxima. The procedure can be summarized as follows:
\begin{enumerate}
    \item The reconstructed energy deposit in each crystal is retained as a hit if it exceeds a noise-related threshold.
    \item Hits satisfying a higher threshold and compatible with being a local maximum with respect to neighbouring crystals are selected as seeds.
    \item For each seed, a core topological cluster is initialized using the seed and its closest neighbours.
    \item The topological cluster is then grown by iteratively aggregating neighbouring hits that are geometrically connected to the current cluster.
    \item If a topological cluster contains multiple seeds, an iterative expectation-maximization procedure, based on a Gaussian-mixture formulation, is used to assign fractions of the hit energies to the individual seed-associated clusters.
\end{enumerate}

Given a topological cluster containing $M$ hits, and denoting by $A_j$ the energy measured in hit $j$ and by $f_{ji}$ the fraction of this energy assigned to the $i^\mathrm{th}$ PFCluster, the reconstructed cluster energy and position can be expressed in the generic form:
\begin{equation}
E_\mathrm{reco}^i = \sum_{j=1}^{M} f_{ji}\, A_j,
\label{eq:pfcluster_energy}
\end{equation}
\begin{equation}
\vec{\mu}_\mathrm{reco}^i =
\frac{1}{E_\mathrm{reco}^i}
\sum_{j=1}^{M} f_{ji}\, A_j\, \vec{c}_j,
\label{eq:pfcluster_position}
\end{equation}

\noindent where $\vec{c}_j$ is the crystal position associated with hit $j$.

\begin{figure*}[ht!]
    \centering
    \begin{subfigure}[t]{0.37\linewidth}
        \centering
        \includegraphics[width=1\linewidth]{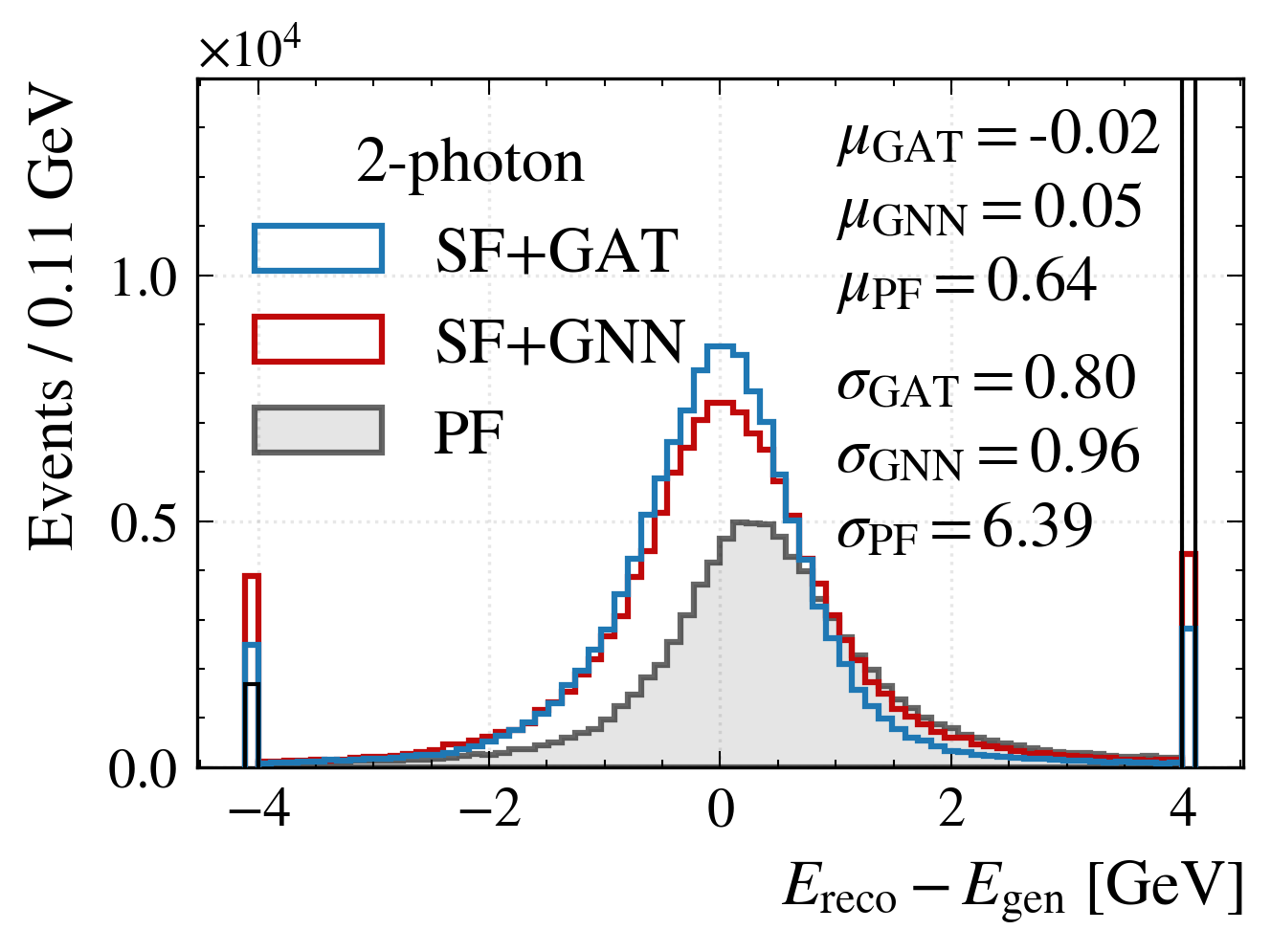}
    \end{subfigure}%
    ~
    \begin{subfigure}[t]{0.63\linewidth}
        \centering
        \includegraphics[width=1\linewidth]{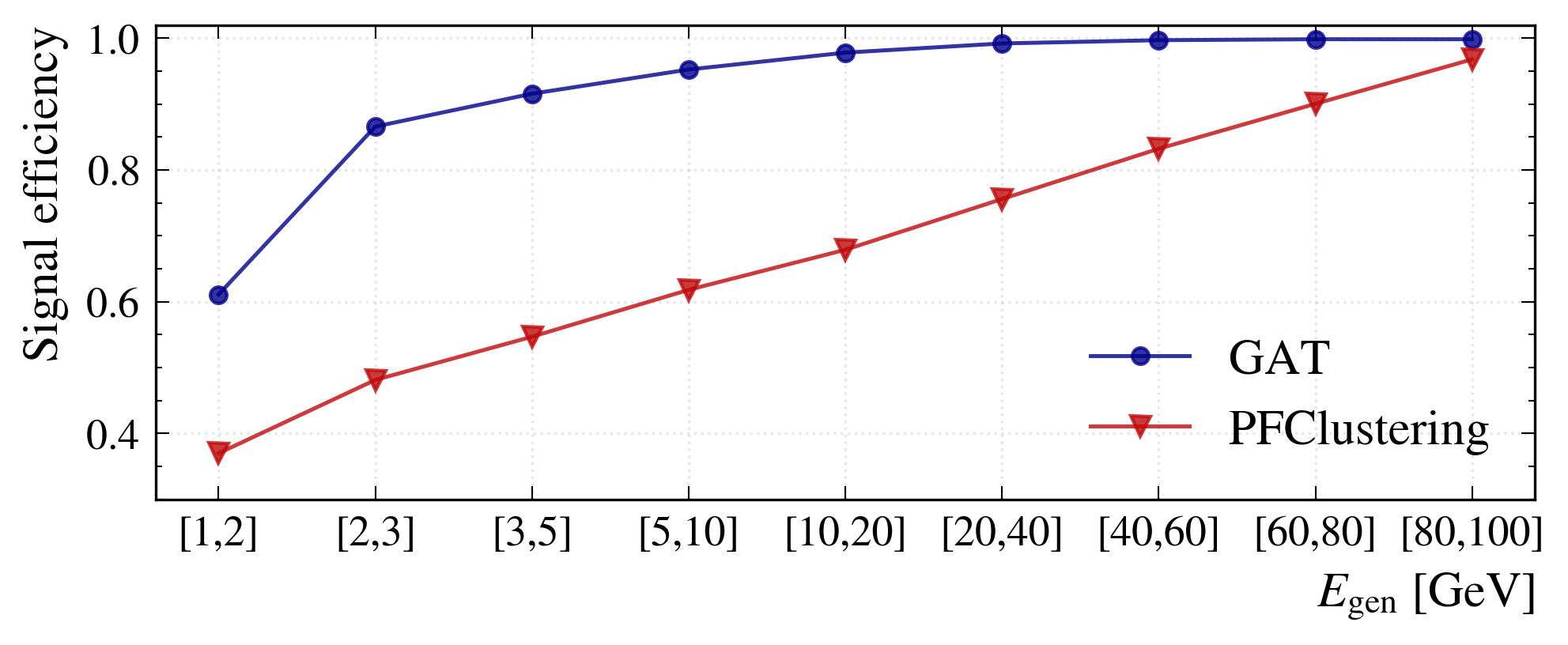}
    \end{subfigure}
    \caption{Toy calorimeter 2-photon testing sample: performance of GAT-$\text{PoEN}_\mathrm{flat}$, GNN-based PoEN and PFClustering~(PF). Left: per-photon energy residuals. Right: per-photon signal efficiency as a function of photon energy. The GAT and GNN working points are chosen to yield consistent signal efficiency}
    \label{fig:pf_perf_2ph}
\end{figure*}

Figure~\ref{fig:pf_perf_2ph} highlights how the overlap-driven limitations of PFClustering appear in practice. The per-photon energy residual distribution in Fig.~\ref{fig:pf_perf_2ph}~(left) shows a markedly broader response for PFClustering in the 2-photon topology, consistent with incorrect energy attribution between nearby showers. The per-photon signal efficiency as a function of photon energy in Fig.~\ref{fig:pf_perf_2ph}~(right) further shows that PFClustering loses efficiency in the low-energy regime, where the second photon is more likely to be absorbed into the dominant cluster or to fail the local-maximum seed requirement. In contrast, the learning-based methods maintain high efficiency and improved energy resolution in the same topology. In particular, the attention-based model retains the resolving power without requiring the ad hoc double-pass procedure used to mitigate spurious reconstructions in the distance-driven message-passing baseline.

\FloatBarrier

\section{Energy and position resolutions}\label{sec:evolution}

\begin{figure*}[ht!]
    \centering
    \begin{subfigure}[t]{0.4\linewidth}
        \centering
        \includegraphics[width=1\linewidth]{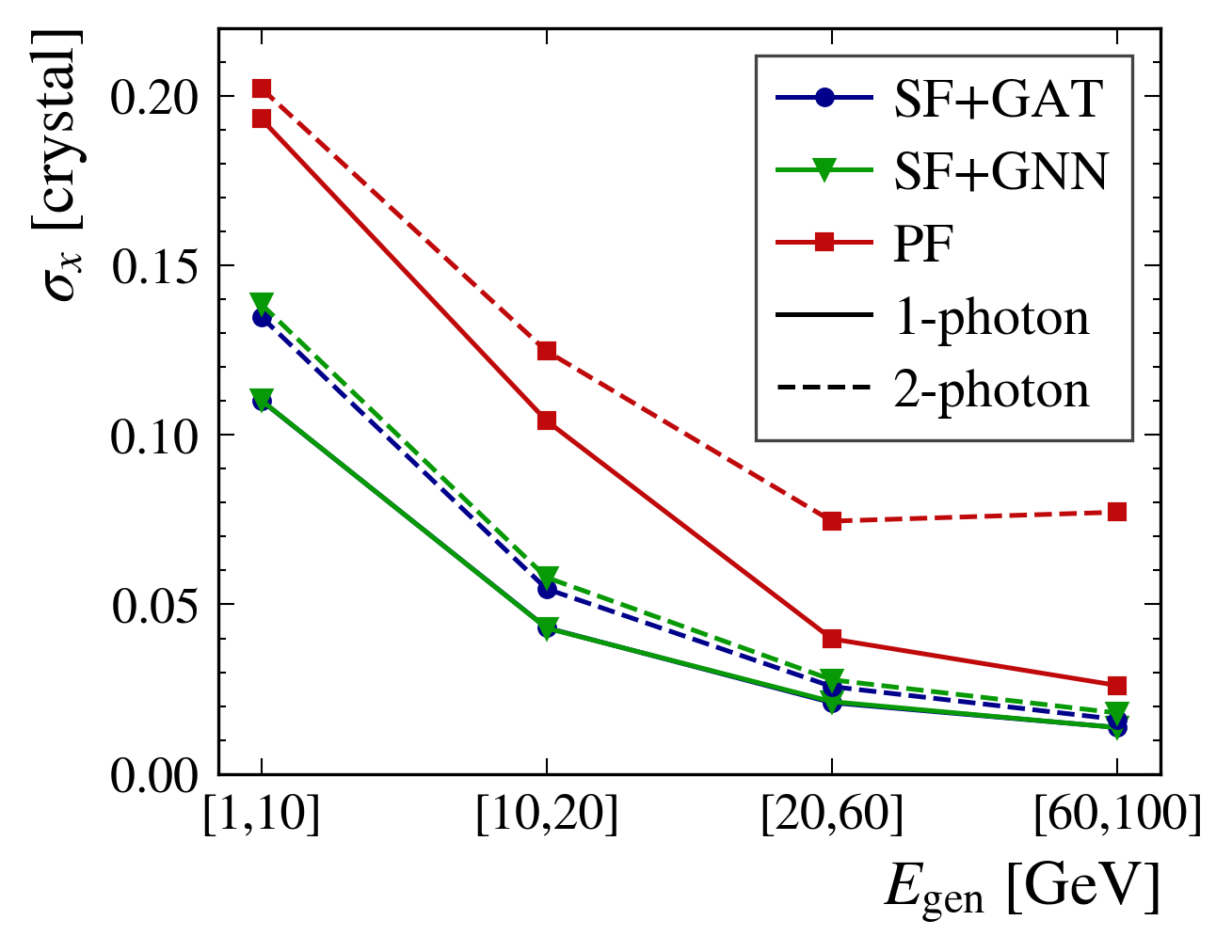}
    \end{subfigure}%
    ~
    \begin{subfigure}[t]{0.4\linewidth}
        \centering
        \includegraphics[width=1\linewidth]{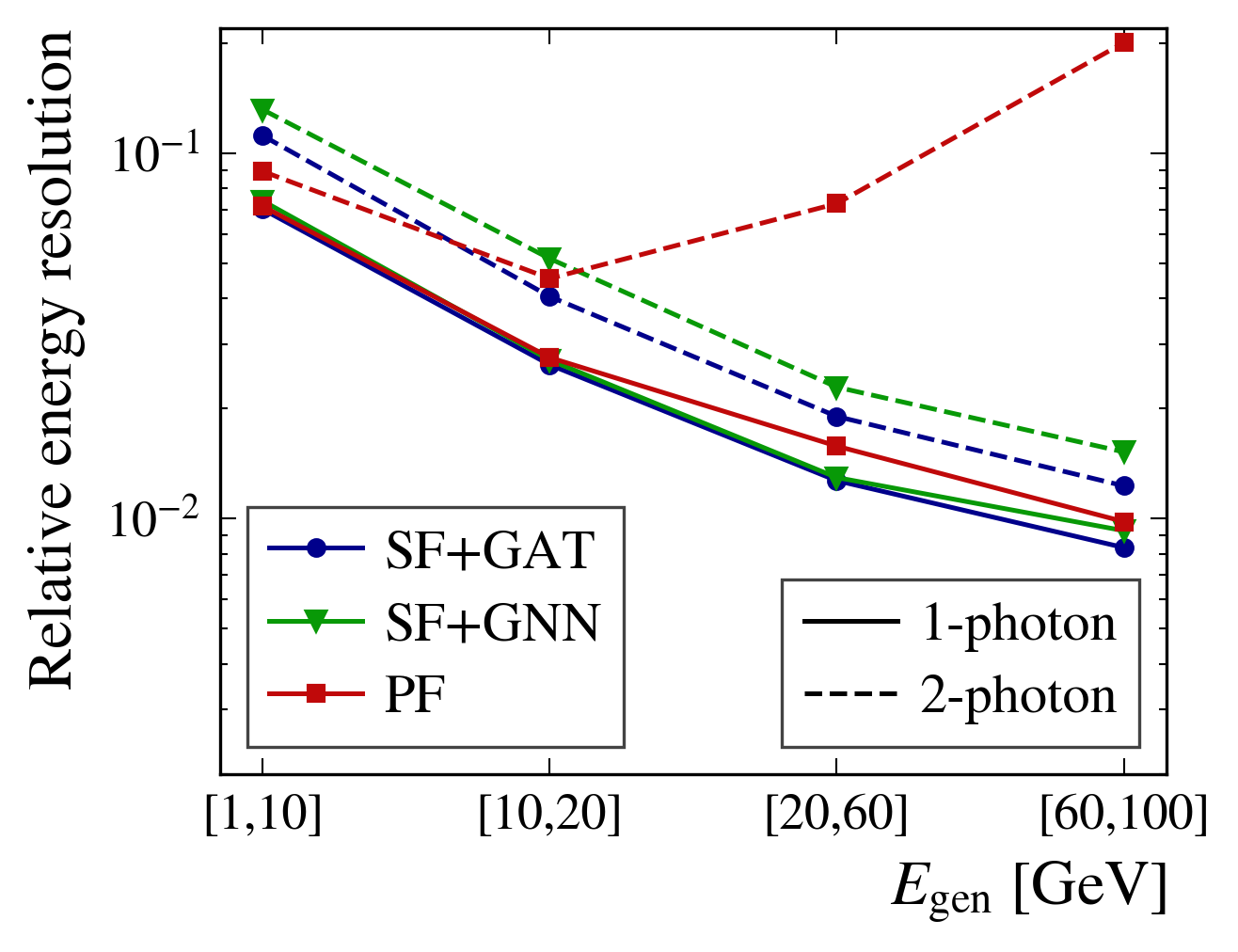}
    \end{subfigure}
    \caption{Toy calorimeter: position resolution (left) and relative energy resolution (right) as a function of $E_\mathrm{gen}$ for the 1-photon and 2-photon topologies. Comparison between GAT-PoEN$_\mathrm{flat}$, GNN-based PoEN and PFClustering (PF)}
    \label{fig:binned_toy}
\end{figure*}

\begin{figure*}[ht!]
    \centering
    \begin{subfigure}[t]{0.4\linewidth}
        \centering
        \includegraphics[width=1\linewidth]{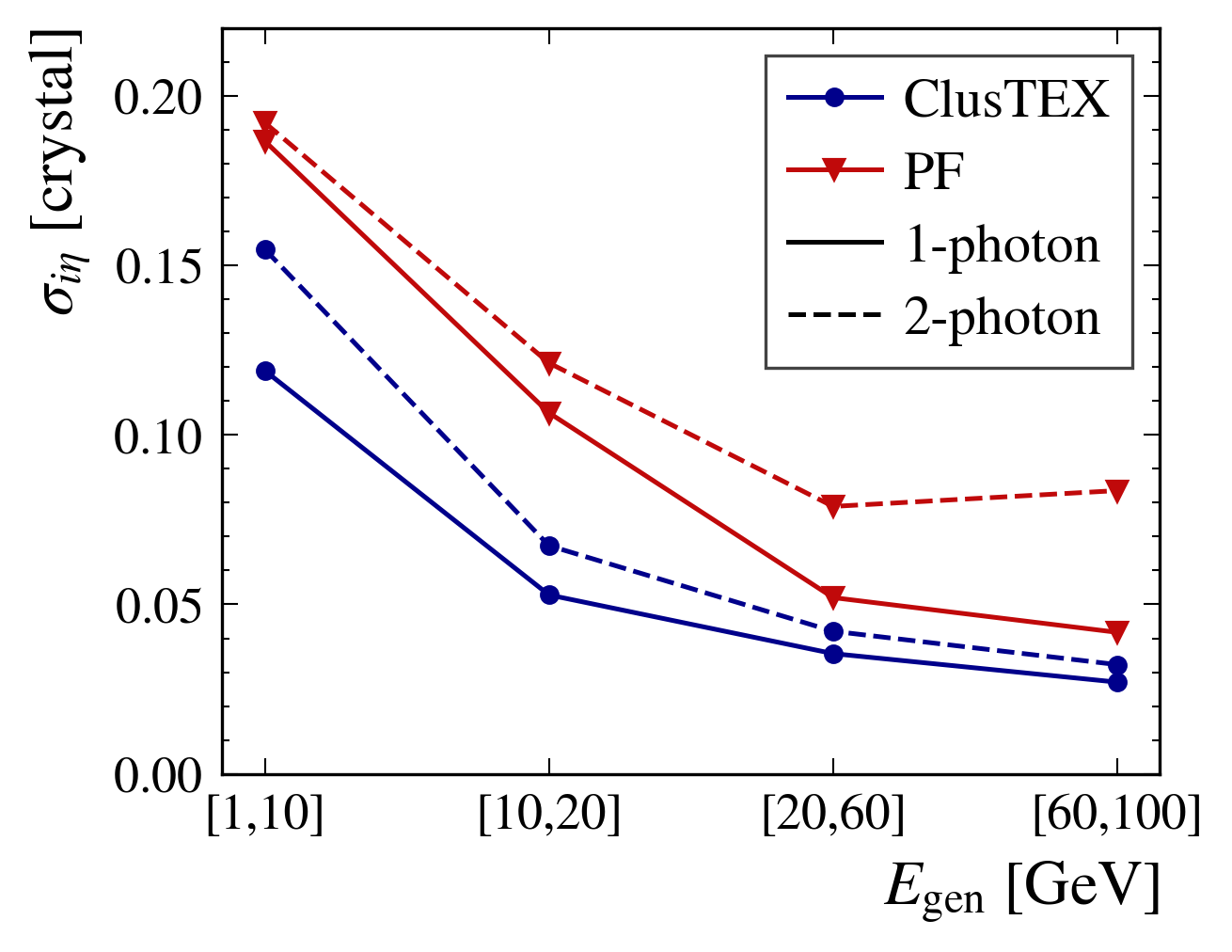}
    \end{subfigure}%
    ~
    \begin{subfigure}[t]{0.4\linewidth}
        \centering
        \includegraphics[width=1\linewidth]{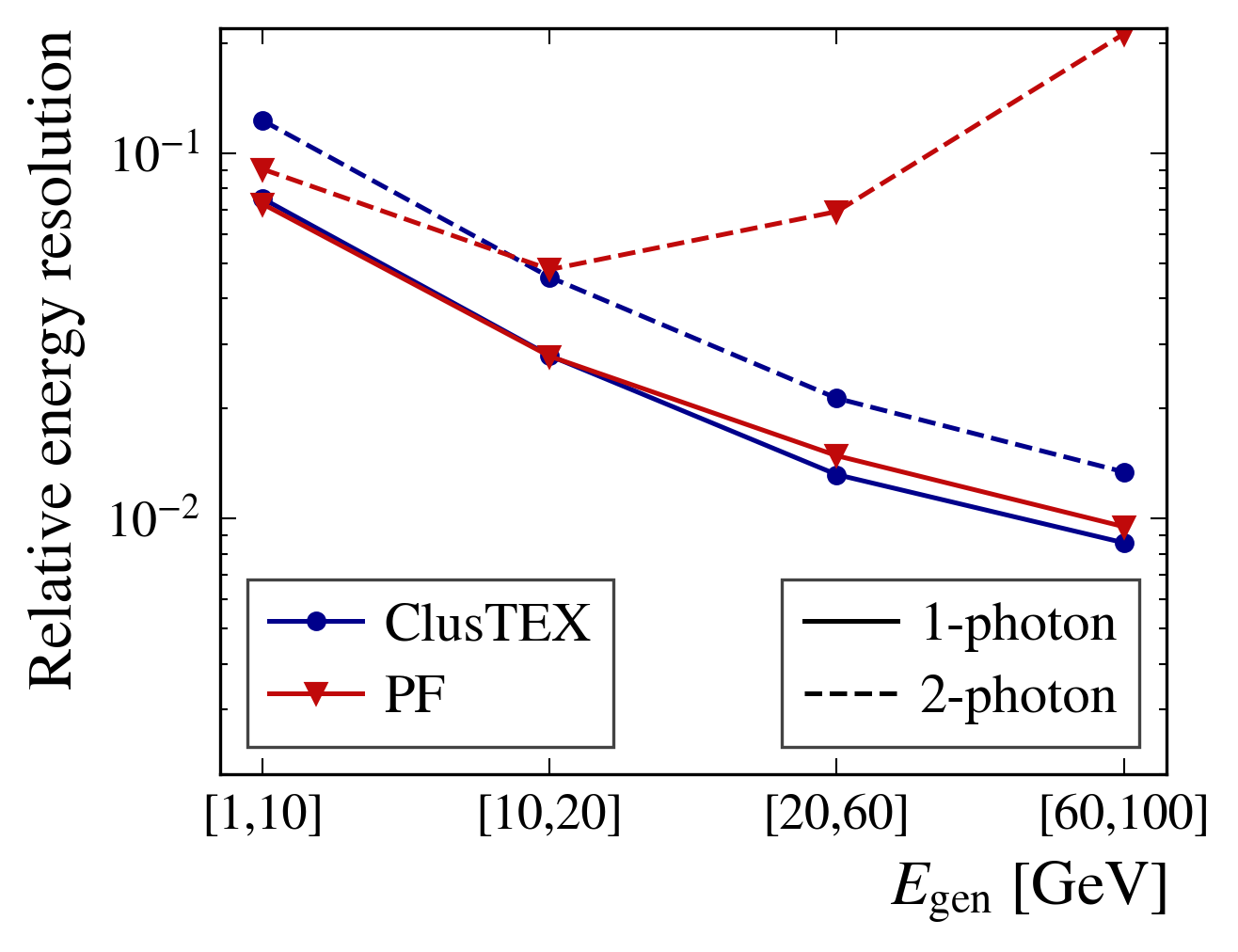}
    \end{subfigure}
    \caption{ECAL-inspired topology: position resolution (left) and relative energy resolution (right) as a function of $E_\mathrm{gen}$ for the 1-photon and 2-photon topologies. Comparison between ClusTEX and PFClustering (PF)}
    \label{fig:binned_ecal}
\end{figure*}

Fig.~\ref{fig:binned_toy} and Fig.~\ref{fig:binned_ecal} show the relative energy resolution $\sigma_E/E_\mathrm{gen}$ and position resolution as a function of the generated photon energy for the toy calorimeter and ECAL-inspired topology respectively. In both cases, the deep learning-based models improve with increasing energy, reflecting the sharper shower profiles and higher signal-to-noise ratio at high $E_\mathrm{gen}$ and achieve better position resolution than PFClustering across all energy bins. The most significant difference appears in the energy resolution of the 2-photon topology where PFClustering's tendency to merge nearby showers  becomes increasingly pronounced at high energies, as discussed in Appendix~\ref{sec:pfclusteringalgo}.

\begin{figure*}[ht!]
    \centering
    \includegraphics[width=0.4\linewidth]{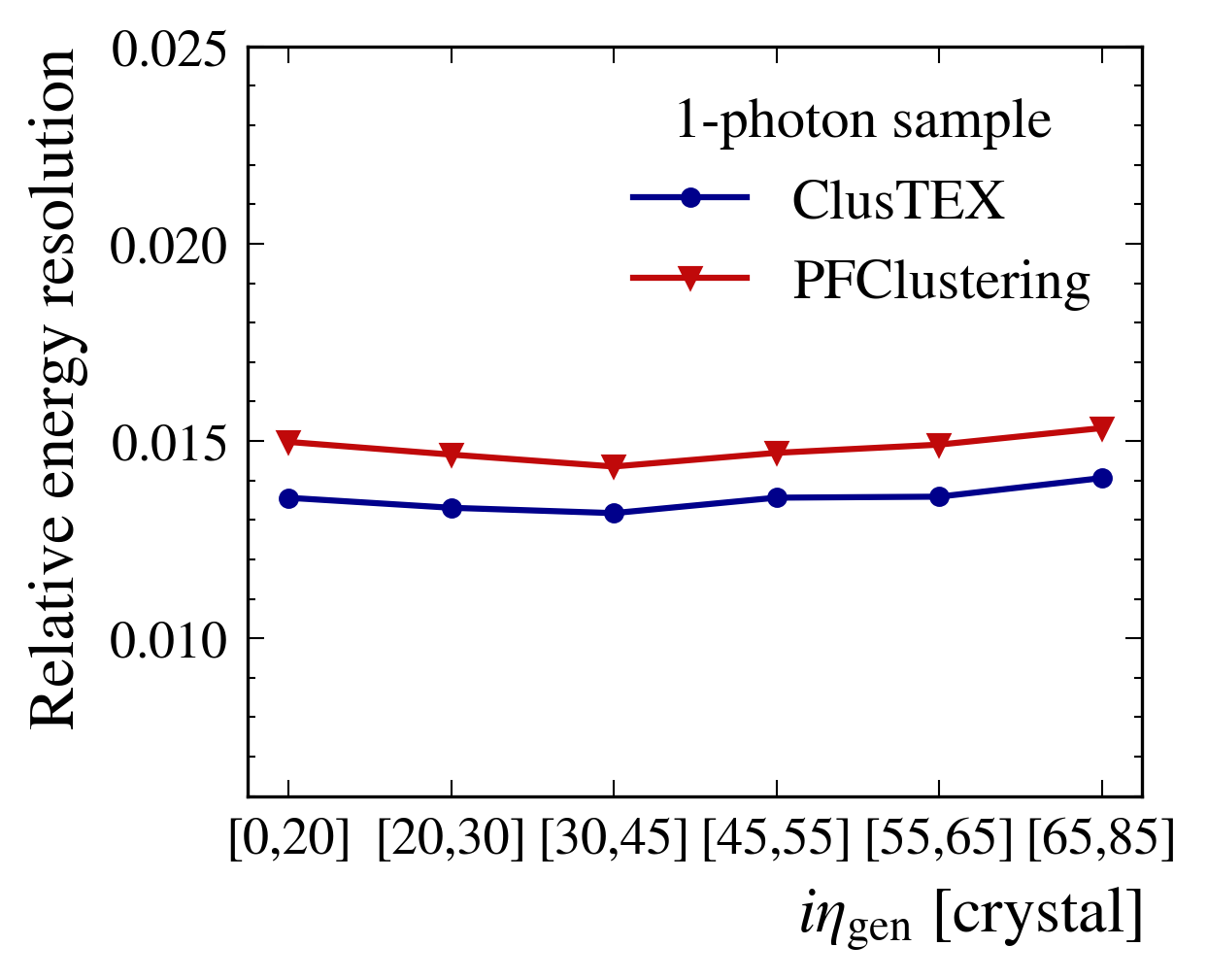}
    \caption{Relative energy resolution as a function of $i\eta_\mathrm{gen}$ for the 1-photon sample as obtained by ClusTEX and PFClustering}
    \label{fig:eta_dependence}
\end{figure*}

The global positional encoding introduced to the ClusTEX architecture handles $\eta$-dependent effects as intended. Fig.~\ref{fig:eta_dependence} shows little dependence of the energy resolution on the photon impact position in the calorimeter.

\FloatBarrier

\section{Graph formation for the graph transformer network}
\label{sec:graph}

Before training and inference, seed objects are defined by iteratively forming local graphs around seed candidates. During inference, the same graph construction is used together with the network predictions to suppress overlaps and reduce duplicates. The procedure is summarized in Algorithm~\ref{alg:graph_form}. In brief, all crystals with deposited energy above a threshold are considered as seed candidates. The highest-energy remaining candidate is selected as an anchor, an overlap window is constructed around it, and all seed candidates inside that overlap window define the nodes of the current graph. For each node, a $7\times7$ input window is extracted and passed to the network. After inference, all candidates in the overlap region are removed from the candidate list and the procedure repeats until no candidates remain.

The overlap window defines the spatial extent over which candidates are processed jointly and therefore sets the maximum graph size. With the choices $d_\mathrm{overlap}=7$ and $d_\mathrm{window}=7$, the resulting local graph covers an effective $13\times13$ neighbourhood around the anchor seed candidate: candidates are selected within a $7\times7$ overlap region, and each candidate contributes a $7\times7$ input window centered on its position. This construction ensures that candidates potentially originating from the same electromagnetic shower, or from overlapping showers, are processed within a common local context, while keeping the graph size bounded and inference tractable.

\begin{algorithm*}[ht!]
\caption{Clustering graph formation}
\label{alg:graph_form}
\vspace{1ex}\textbf{Parameters tuned on simulation:}\\[1ex]
{\small
\quad $E_\mathrm{dep}^{\mathrm{thresh}} = 0.5 \,\text{GeV}$\hspace{2ex}
\quad $d_\mathrm{overlap} = 7$\hspace{2ex}
\quad $d_\mathrm{window} = 7$ \hspace{2ex}
\quad $d_\mathrm{effective} = d_\mathrm{overlap} + d_\mathrm{window} - 1$ \\[1ex]
}
\textbf{Initialisation:}\\[0.5ex]
{\small
Select all cells with deposited energy above $E_\mathrm{dep}^{\mathrm{thresh}}$ as seed candidates. Order them in descending order of deposited energy.\\[1ex]
}
\textbf{Iterative procedure while seed candidates remain:}\\[0.5ex]
{\small
\textbf{Step 1:} Select the highest-$E_\mathrm{dep}$ remaining seed candidate and define it as the anchor.\\
\textbf{Step 2:} Construct an overlap window of width $d_\mathrm{overlap}$ centered on the anchor.\\
\textbf{Step 3:} Identify all seed candidates contained within this overlap window. For each identified candidate, construct an input window of width $d_\mathrm{window}$ centered on the candidate.\\
\textbf{Step 4:} Pass all input windows to the network to obtain the seed probability, relative position $(x_\mathrm{reco},y_\mathrm{reco})$, and reconstructed energy $E_\mathrm{reco}$.\\
\textbf{Step 5:} Eliminate all seed candidates contained within the overlap window from the candidate list.\\
}
\end{algorithm*}

\begin{figure}
    \centering
    \includegraphics[width=0.7\linewidth]{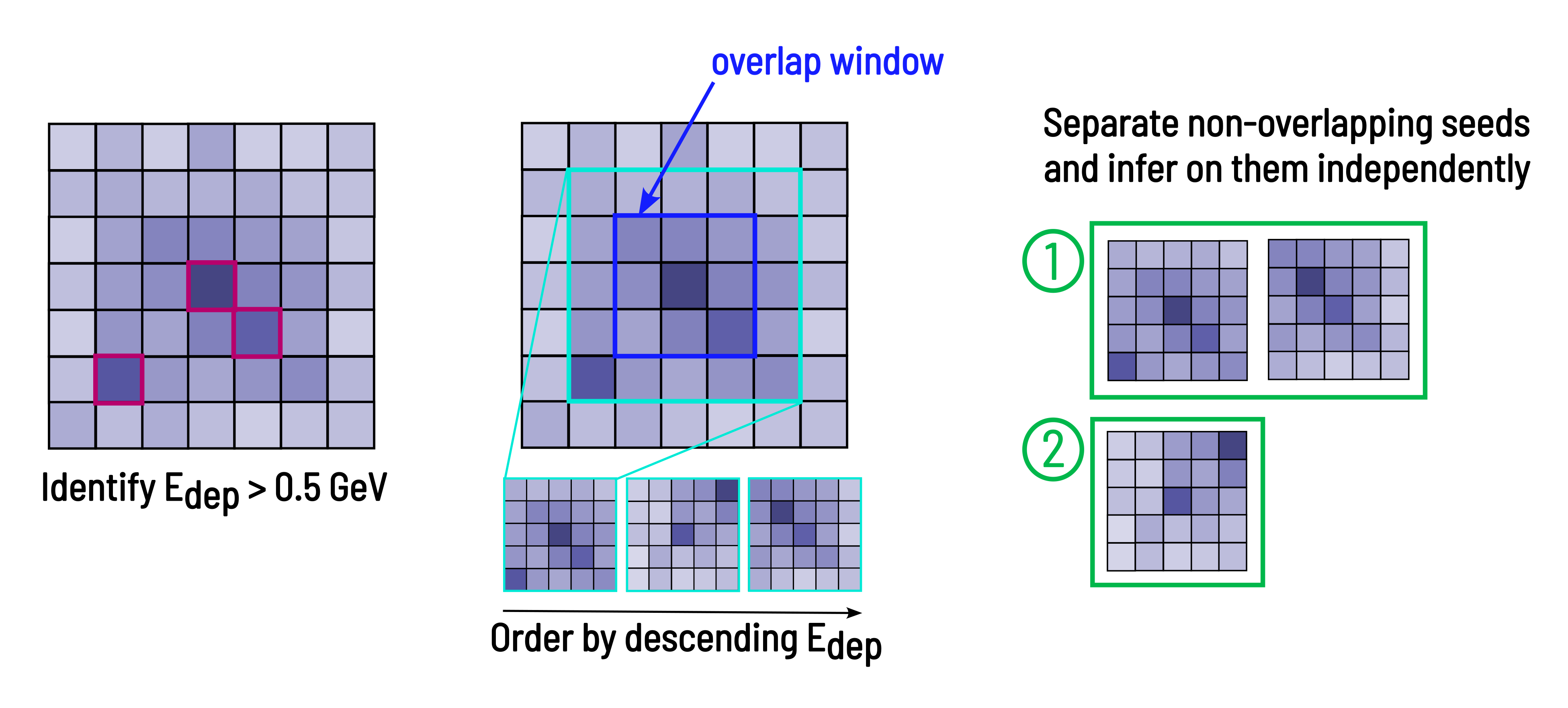}
    \caption{Schematic illustration of the graph formation procedure. Seed candidates above threshold are iteratively clustered into local graphs around the highest-energy anchor candidate using an overlap window of width $d_\mathrm{overlap}$. The illustration is scaled down to $d_\mathrm{window}=5$ and $d_\mathrm{overlap}=3$}
    \label{fig:graphing}
\end{figure}

\begin{figure}[H]
    \centering
    \includegraphics[width=0.9\linewidth]{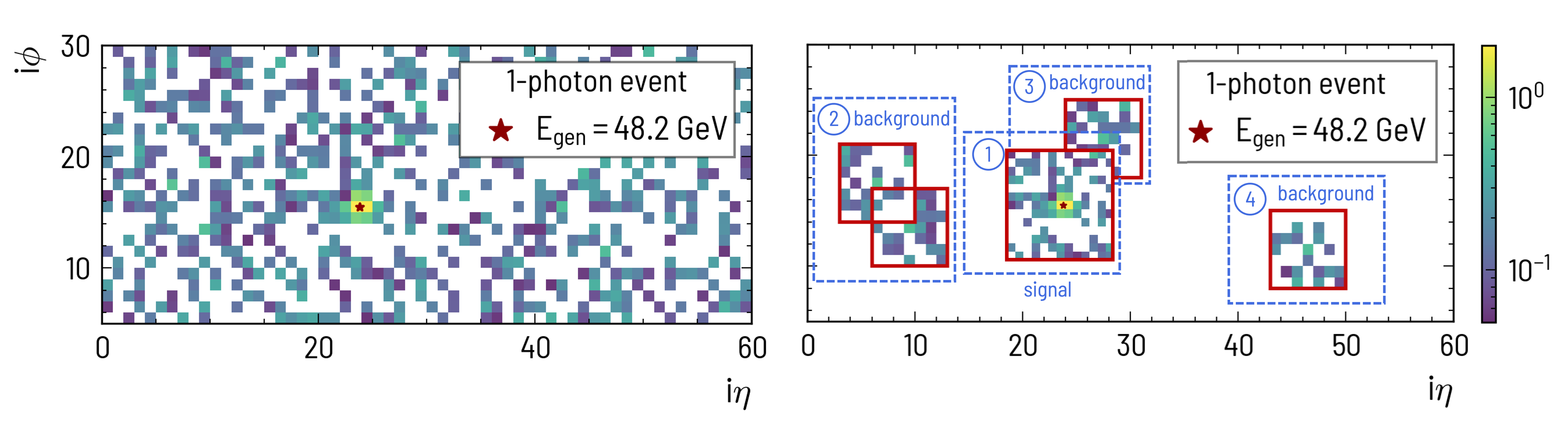}
    \caption{Example calorimeter window illustrating candidate selection and graph construction. Left: simulated energy deposits including electronic noise. Right: seed candidates within the overlap region form the nodes of one local graph, yielding one signal-centered window (label 1) and additional background-centered candidate windows (labels 2--4)}
    \label{fig:neighbourhood}
\end{figure}

\end{appendices}

\bibliography{sn-bibliography}% common bib file

@article{exotic_higgs,
    author="{CMS Collaboration}",
    title="{“Search for exotic Higgs  boson decays H $\to$ $\mathcal{A} \mathcal{A}$ $\to$ 4$\gamma$ with event containing two merged diphotons in proton-proton collisions at $\sqrt{s}$ = 13 TeV"}",
    journal = "Phys. Rev. Lett.",
    volume = "131",
    number = "10",
    pages = "101801",
    year = "2023",
    doi = {10.1103/PhysRevLett.131.101801}
}

@article{cms,
author ="{CMS Collaboration}", 
title = "{“The CMS experiment at the CERN LHC”}",
journal = {Journal of Instrumentation},
doi = {10.1088/1748-0221/3/08/S08004},
url = {https://dx.doi.org/10.1088/1748-0221/3/08/S08004},
year = {2008},
month = {aug},
publisher = {},
volume = {3},
number = {08},
pages = {S08004}
}

@article{CMS:2020uim,
    author = "{CMS Collaboration}",
    title = "{“Electron and photon reconstruction and identification with the CMS experiment at the CERN LHC”}",
    reportNumber = "CMS-EGM-17-001, CERN-EP-2020-219",
    journal = "JINST",
    volume = "16",
    number = "05",
    pages = "P05014",
    year = "2021",
    doi = {10.1088/1748-0221/16/05/P05014},  
    url = "https://doi.org/10.1088/1748-0221/16/05/P05014"
}

@article{higgs,
    author = "{CMS Collaboration}",
    title = "{“Measurements of the Higgs boson production cross section and couplings in the W boson pair decay channel in proton-proton collisions at $\sqrt{s}=13~\text{TeV}$”}",
    reportNumber = "CMS-HIG-20-013, CERN-EP-2022-120",
    doi = "10.1140/epjc/s10052-023-11632-6",
    journal = "Eur. Phys. J. C",
    volume = "83",
    number = "7",
    pages = "667",
    year = "2023"
}

@article{art1,
    author = "{Belayneh, D. et al.}",
    title = "{“Calorimetry with deep learning: particle simulation and reconstruction for collider physics”}",
    doi = {10.1140/epjc/s10052-020-8251-9},
    journal = "Eur. Phys. J. C",
    volume = "80",
    number = "7",
    pages = "688",
    year = "2020"
}

@article{art2,
    author = "{Akchurin, N. ; Cowden, C. ; Damgov, J. ; Hussain, A. ; Kunori, S.}",
    title = "{“On the use of neural networks for energy reconstruction in high-granularity calorimeters”}",
    doi = {10.1088/1748-0221/16/12/P12036},
    journal = "JINST",
    volume = "16",
    number = "12",
    pages = "P12036",
    year = "2021"
}

@article{calorimetry,
     author = "{Fabjan, C. W. and Gianotti, F.}",
      title        = "{“Calorimetry for Particle Physics”}",
      reportNumber  = "CERN-EP-2003-075",
      doi ={10.1103/RevModPhys.75.1243},
      journal       = "Rev. Mod. Phys.",
      volume        = "75",
      pages         = "1243-1286",
      year          = "2003",
      url           = "https://cds.cern.ch/record/692252",
}

@article{pfclustering,
    author = "{CMS Collaboration}",
    title = "{“Particle-flow reconstruction and global event description with the CMS detector”}",
    journal = {Journal of Instrumentation},
    doi= {10.1088/1748-0221/12/10/P10003},
    url = {https://dx.doi.org/10.1088/1748-0221/12/10/P10003},
    year = {2017},
    month = {oct},
    publisher = {},
    volume = {12},
    number = {10},
    pages = {P10003},
}

@article{geant,
    author = "{Agostinelli, S. et al.}",  
    title = "{“Geant4—a simulation toolkit"}",
    journal = {Nuclear Instruments and Methods in Physics Research Section A: Accelerators, Spectrometers, Detectors and Associated Equipment},
    volume = {506},
    number = {3},
    pages = {250-303},
    year = {2003},
    doi = {10.1016/S0168-9002(03)01368-8}
}

@techreport{cms_tdr,
    author = "{CMS Collaboration}",
    title = "{“The CMS electromagnetic calorimeter project: Technical Design Report"}",
    reportNumber = "CERN-LHCC-97-033, CMS-TDR-4, CERN-LHCC-97-033, CMS-TDR-4",
    year = "1997", 
    url = "https://cds.cern.ch/record/349375?ln=en",
    institution = {CERN}
}

@inbook{BDT,
   chapter={“Boosted Decision Trees"},
   doi={10.1142/9789811234033_0002},
   title={Artificial Intelligence for High Energy Physics},
   publisher={World Scientific},
   author="{Coadou, Y.}",
   year= {2022},
   address = {ISBN: 978-981-12-3403-3},
   pages= {9–58} }

@misc{ecal_constant,
      author = "{CMS Collaboration}", 
      title         = {“ECAL 2016 refined calibration and Run2 summary plots"},
      year          = "2020",
      url           = "https://cds.cern.ch/record/2717925"
}

@article{higgs_mass,
title = "{“A measurement of the Higgs boson mass in the diphoton decay channel”}",
author = "{CMS Collaboration}", 
journal = {Physics Letters B},
volume = {805},
pages = {135425},
year = {2020},
issn = {0370-2693},
doi = {https://doi.org/10.1016/j.physletb.2020.135425},
url = {https://www.sciencedirect.com/science/article/pii/S037026932030229X}
}

@misc{clusterin_run3,
      author = "{CMS Collaboration}", 
      title         = "{“ECAL Clustering for run 3”}",
      year          = "2022",
      url           = "https://cds.cern.ch/record/2812783"
}

@article{dcluster,
    author = "{Simkina, P. and Couderc, F. and Malclès, J. and Sahin, M. Ö.}",
    title = "{“Reconstruction of electromagnetic showers in calorimeters using Deep Learning”}",
    doi = "10.1140/epjc/s10052-024-12978-1",
    journal = "Eur. Phys. J. C",
    volume = "84",
    number = "6",
    pages = "639",
    year = "2024"
}

@inproceedings{tbmatrix,
  author    = {T. Frisson and P. Mine},
  title     = {H4SIM, a Geant4 simulation program for the CMS ECAL supermodule},
  booktitle = {Geant4 10th International Conference},
  year      = {2005},
  month     = nov,
  address   = {Bordeaux, France},
  url       = {http://geant4.in2p3.fr/2005/Workshop/UserSession/P.Mine.pdf}
}

@phdthesis{simkina:tel-04412128,
  TITLE = {{Event reconstruction and analysis in CMS using artificial intelligence}},
  AUTHOR = {Simkina, Polina},
  URL = {https://theses.hal.science/tel-04412128},
  NUMBER = {2023UPASP095},
  SCHOOL = {{Universit{\'e} Paris-Saclay}},
  YEAR = {2023},
  MONTH = Sep,
  TYPE = {Theses},
  HAL_ID = {tel-04412128},
  HAL_VERSION = {v1},
}

@article{Valsecchi_2023,
doi = {10.1088/1742-6596/2438/1/012077},
url = {https://doi.org/10.1088/1742-6596/2438/1/012077},
year = {2023},
month = {feb},
publisher = {IOP Publishing},
volume = {2438},
number = {1},
pages = {012077},
author = {Valsecchi, Davide and for the CMS Collaboration},
title = "{“Deep learning techniques for energy clustering in the CMS ECAL”}",
journal = {Journal of Physics: Conference Series},
}

@Article{art3,
AUTHOR = {Schmidt, Kylian and Kota, Krishna Nikhil and Kieseler, Jan and De Vita, Andrea and Klute, Markus and Abhishek and Aehle, Max and Awais, Muhammad and Breccia, Alessandro and Carroccio, Riccardo and Chen, Long and Dorigo, Tommaso and Gauger, Nicolas R. and Lupi, Enrico and Nardi, Federico and Nguyen, Xuan Tung and Sandin, Fredrik and Willmore, Joseph and Vischia, Pietro},
TITLE = "{“End-to-End Detector Optimization with Diffusion Models: A Case Study in Sampling Calorimeters”}",
JOURNAL = {Particles},
VOLUME = {8},
YEAR = {2025},
NUMBER = {2},
ARTICLE-NUMBER = {47},
URL = {https://www.mdpi.com/2571-712X/8/2/47},
ISSN = {2571-712X},

DOI = {10.3390/particles8020047}
}

@article{art4,
doi = {10.1088/2632-2153/ae5c56},
url = {https://doi.org/10.1088/2632-2153/ae5c56},
year = {2026},
month = {apr},
publisher = {IOP Publishing},
volume = {7},
number = {2},
pages = {025061},
author = {Nguyen, Xuan Tung and Chen, Long and Dorigo, Tommaso and Gauger, Nicolas R and Vischia, Pietro and Nardi, Federico and Awais, Muhammad and Hanif, Hamza and Abbas, Shahzaib and Kapoor, Rukshak},
title = "{“Differentiable surrogate for detector simulation and design with diffusion models”}",
journal = {Machine Learning: Science and Technology},
}

@dataset{zenodo,
  author       = "{Maidannyk, Y. and Sahin, M. Ö.}",
  title        = {GEANT4 Simulation Dataset for a CMS-Like PbWO4 Electromagnetic Calorimeter},
  month        = mar,
  year         = 2026,
  publisher    = {Zenodo},
  version      = {0.0.1},
  doi          = {10.5281/zenodo.18929909},
}
%% if required, the content of .bbl file can be included here once bbl is generated
%%\input sn-article.bbl

\end{document}